\documentclass[trackchanges, twocolumn]{aastex701}
\usepackage{hyperref}
\hypersetup{
    colorlinks=true,
    linkcolor=blue,
    filecolor=blue,      
    urlcolor=blue,
    }
\usepackage{soul}
\usepackage[normalem]{ulem}
\usepackage{epstopdf}
\usepackage{float}
\usepackage{graphicx}
\usepackage{subcaption} 

\shorttitle{Dynamics of Interstellar Comet 3I/ATLAS}
\shortauthors{G. Ahuja \& S. Ganesh}
\submitjournal{ApJL}

\begin{document}

\title{Dynamical simulation of the Interstellar Comet 3I/ATLAS}

\correspondingauthor{Goldy Ahuja}
\author[orcid=0009-0008-1809-3256,gname='Goldy',sname='Ahuja']{Goldy Ahuja}
\affiliation{Physical Research Laboratory, Ahmedabad, Gujarat-380009, India}
\affiliation{Indian Institute of Technology Gandhinagar, Palaj, Gujarat-382355, India}
\email[show]{goldy@prl.res.in, goldezz20@gmail.com}  

\author[orcid=0000-0002-7721-3827,gname=Shashikiran, sname=Ganesh]{Shashikiran Ganesh} 
\affiliation{Physical Research Laboratory, Ahmedabad, Gujarat-380009, India}
\email{shashi@prl.res.in}

\begin{abstract}


Comet 3I/ATLAS, also known as C/2025 N1, was discovered on 2025 July 1 UT by NASA Asteroid Terrestrial-impact Last Alert System (ATLAS), with a v$_{\infty}$ $\sim$ 58 kms$^{-1}$. This is the fastest among the three interstellar objects discovered so far. In this work, we study the interaction of the 3I/ATLAS with Mars, pre-perihelion, and Jupiter post-perihelion. 
We also present the results of the dynamical simulations of the orbital evolution of the comet for a hundred years in the past and future. 
For our analysis, we have used REBOUND, an N-Body simulation package, to study these situations. We use the adaptive size mathematical integrator \textsc{Ias15}, with a 1-day time step for long-term integration, and a 1-hour time step to study the effect of the planets on this body during the close encounters. We have seen an effect of Jupiter on the orbital parameters of the comet, which affects its post-perijove trajectory significantly. The impact of Mars on this comet is minimal compared to the effect of Jupiter. This is consistent with the point that the comet moves well past Mars's Hill radius but very close to Jupiter's Hill radius at the respective close approaches. However, the effect of non-gravitational forces will alter the results. Since the non-gravitational forces are not yet known, we predict the variation of the orbital parameters considering a range of possible magnitudes of the non-gravitational acceleration. 
\end{abstract}



\section{Introduction}
Interstellar comets are small bodies that are formed outside our solar system and are found passing through the Solar system. They are expected to have been formed in a protoplanetary disk and then thrown out of the parent star system, possibly due to gravitational scattering by heavy bodies, such as planets, present in the disk \citep{2023ARA&A..61..197J}. They travel through the ISM before encountering any other stellar system, such as the solar system. Two interstellar visitors have been recorded passing through the solar system. The first one was 1I/$'$Oumuamua or C/2017 U1, a metallic asteroid-like object with no visible coma \citep{Meech_1I_2017,Quan_1I_2017,Oumuamua_ISSI_2019}. The second one was the comet 2I/Borisov or C/2019 Q4, well studied by many authors \citep{Opitom_2I_2019,Fitzsimmons_2I_2019, Jewitt_2I_2020, Piotr_2I_Nature2020,Bodewits_2I_2020,Aravind_2I_2021, prodan_2I_2024}.  Recently, the third interstellar object, 3I/ATLAS or C/2025 N1, hereafter 3I, has been discovered at a heliocentric distance of 4.51 au, with an eccentricity of $6.13 \pm 0.02$ \citep{3I_disc_1,3I_disc_2}. The comet is moving with a very high radial velocity of around 58 kms$^{-1}$ and was at a perihelion distance of $1.357 \pm 0.001$ au in October 2025. 

\begin{figure*}
\plottwo{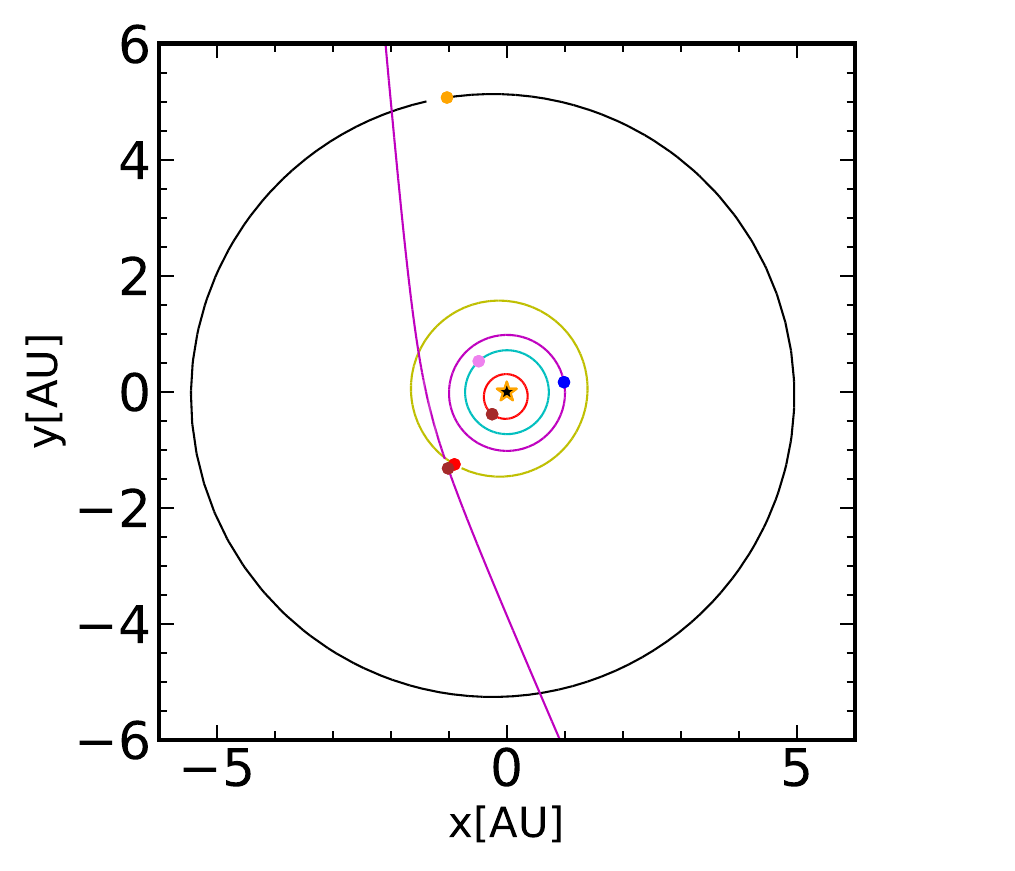}{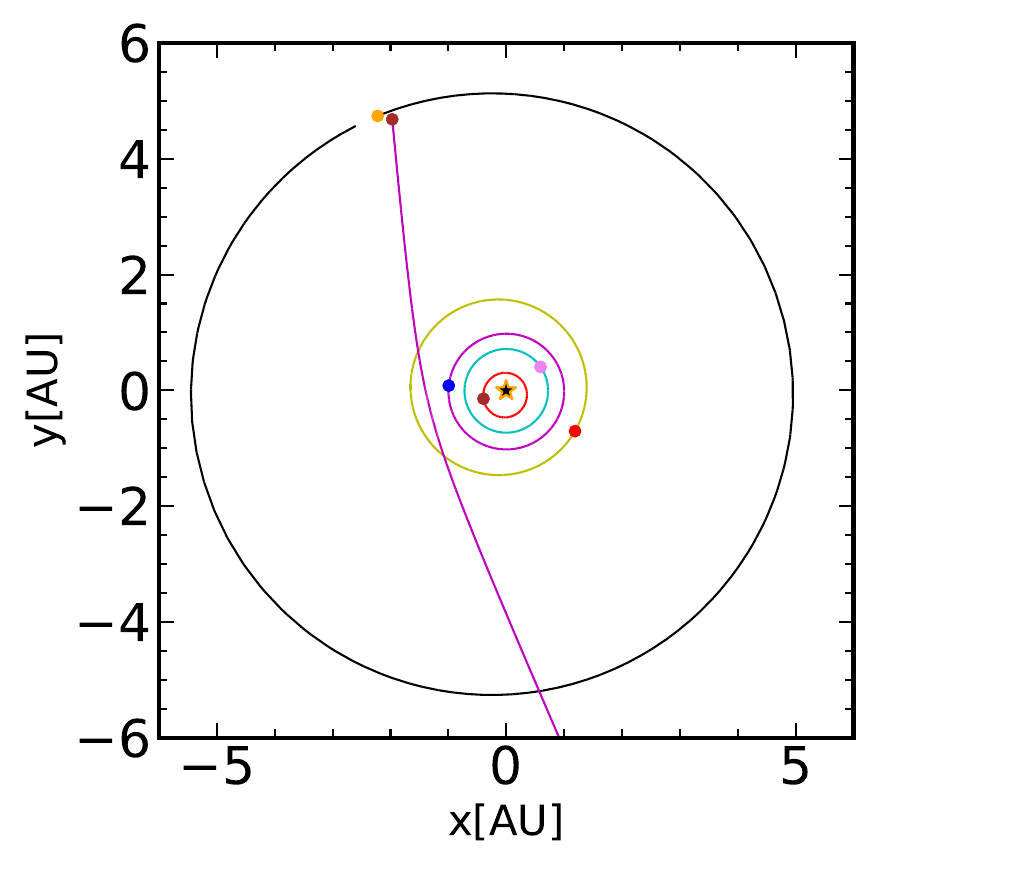}
\caption{Left: Close encounters between interstellar comet 3I/ATLAS with Mars at Epoch JD 2460951.5 (2025 October 03). Right: Close encounters between interstellar comet 3I/ATLAS with Jupiter at Epoch JD 2461115.5 (2026 March 16). For clarity, we show only the planets' orbits(up to Jupiter), whereas, in the simulation, we have also considered the three other planets (Saturn, Uranus, Neptune), the four big asteroids (Ceres, Pallas, Juno, Vesta) in the asteroid belt, and five KBOs (Pluto, Makemake, Haumea, Quaoar, Sedna).}
\label{fig: orbits}
\end{figure*}

The interstellar comet 3I made a close approach to Mars at a distance of 0.194 au on Epoch JD 2460951.5 (2025 October 03)\footnote{\href{https://ssd.jpl.nasa.gov/horizons_batch.cgi?batch=1&COMMAND=\%273I\%27&CENTER=\%27500@499\%27&START_TIME=\%272025-10-03\%2002:00\%27&STOP_TIME=\%272025-10-03\%2007:00\%27&STEP_SIZE=\%2730\%20min\%27&QUANTITIES=\%2720,39\%27}{HORIZONS ephemeris for interaction with Mars}} and will be approaching Jupiter at a distance of 0.357 au on Epoch JD 2461115.5 (2026 March 16)\footnote{\href{https://ssd.jpl.nasa.gov/tools/sbdb_lookup.html\#/?sstr=3I\%2FATLAS}{NASA SBDB lookup showing close approach with Jupiter}}. The interaction of the comet with Mars and Jupiter is shown in the figure \ref{fig: orbits}. At those distances from the respective planets, the comet's orbital elements will be affected due to gravitational perturbation from these planets, resulting in a change in trajectory.

Alongside the gravitational perturbations, comets are also subject to non-gravitational forces due to the sublimation of volatile ices, which exerts a recoil acceleration on the comet's nucleus, affecting their orbital evolution \citep{Marsden_1973,Hui_2017}. For example, both the previously seen interstellar objects have been subjected to non-gravitational forces \citep{Oumuamua_nongrav_2018, 2023ARA&A..61..197J}.

The evolution of a comet's orbit is studied numerically using dynamical N-body simulation codes. Such dynamical simulations \citep{Fuente_2012_2002VE68,Królikowska_2010_250au, Dynczynski_2011_250au,Fuente_2014_ND15,Gab_2024} have become a viable option in recent years to study the long-term effects of gravitational and non-gravitational forces and to examine any close encounters with the planets \citep{Shober_2024,Anderson_2024,Pilorz_2025}.

In this letter, we discuss the dynamics of interstellar comet 3I at different timescales. We start our discussion with the past (and future) dynamical trajectory by integrating backwards (and forwards) for timescales of 100 years, with a timestep of 1 day. We study the effect of the interactions with Mars and Jupiter at a timestep of 1 hour and limit our simulation to a 20-year period (at which point the comet reaches close to 250 au - just outside the planetary perturbation limits \citep{Królikowska_2010_250au,Dynczynski_2011_250au}).

\begin{deluxetable*}{lccc}
\tablewidth{100pt}
\tablecaption{Heliocentric and Barycentric Orbital Elements of Interstellar comet 3I/ATLAS. \label{tab: orbit}}
\tablehead{
\colhead{Orbital Parameters} & \colhead{Units} & \colhead{Heliocentric} & \colhead{Barycentric}
}
\startdata
Perihelion Distance (q) & au & 1.3572 $\pm$ 0.0006 & 1.3626\\
Semi-major axis (a) & au &  - 0.2638 $\pm$ 0.0001 & -0.2642\\
Eccentricity (e) & -- & 6.145 $\pm$ 0.004 & 6.157\\
Inclination (i) & degree & 175.1135 $\pm$ 0.0003 & 175.1275 \\
Longitude of Ascending Node ($\Omega$) & degree & 322.166 $\pm$ 0.006 & 322.255\\
Argument of Perihelion ($\omega$) & degree & 128.004 $\pm$ 0.005 & 128.083\\
\enddata
\tablecomments{The Epoch of these elements is JD 2460867.5 (2025-07-11 00:00:00). The results are based on 601 observations used, which covers the data span of 58 days\textsuperscript{\ref{myfn}}.}
\end{deluxetable*}

In section \ref{sec: Methods}, we discuss the methodology followed to set up the dynamical simulations and compute the interaction of comet 3I with Mars and Jupiter. In section \ref{sec: Results}, we show the past and future orbital evolution of this comet for 100 years under the effect of gravitational perturbation.
We also study the effect of the interactions with Mars and Jupiter by considering two different simulations - one with both planets and one in their absence.
In section \ref{sec: NGF}, we assume a range of non-gravitational accelerations that the comet may be subjected to, and predict the variation of the orbital parameters under the effect of both gravitational and non-gravitational forces.

\section{Methodology} \label{sec: Methods}
To understand comet 3I's long-term past and future dynamical behavior, we have used the \textsc{Python}-based REBOUND N-body simulation package \citep{rebound}.
We have used different integrators to solve for the two timescales as discussed below. For long-term integrations in the past and future to see the projection of the comet, and to study the interaction with planets such as Mars and Jupiter, we have used the non-symplectic mathematical integrator, i.e., Integrator with Adaptive Step-size control, 15th order (\textsc{Ias15}), which is very accurate up to machine precision \citep{reboundias15}.
For long-term integration, the time step of 1 day is chosen, and for the planetary interaction, we have considered 1 hour to be optimal to study changes in the different orbital parameters and the comet's Cartesian velocities.

We make use of the NASA JPL's Small Body Database lookup (SBDB)\footnote{\label{myfn}\url{https://ssd.jpl.nasa.gov/tools/sbdb_lookup.html\#/?sstr=3I\%2FATLAS}} and JPL Horizons service\footnote{\url{https://ssd.jpl.nasa.gov/horizons/}}\citep{JPL} to get the orbital elements and covariance matrices of the orbital elements of the comet 3I at Epoch JD 2460867.5 (2025 July 11, 00:00:00 UT). The Heliocentric orbital elements with 1-$\sigma$ uncertainties and the corresponding Barycentric orbital elements of comet 3I are given in Table \ref{tab: orbit}. After getting the covariance matrix, we followed the approach of \citet{Egal_multivar_2022,sarah_multi_2023} to create 500 massless clones of the interstellar comet 3I using the multivariate distribution.

The orbits of the statistical clones are numerically integrated in the presence of the eight major planets, four of the large asteroids present in the asteroid belt, i.e, Ceres, Pallas, Juno, and Vesta, and the five big KBOs, namely, Pluto, Makemake, Haumea, Quaoar, and Sedna.
We consider the barycenter of the planets at a similar epoch to the comet 3I in the simulation.
The values and the error bars of the different quantities, reported in the text, are calculated by taking the mean and the standard deviation of the values of the parameters for all 500 clones. 

\section{Results and discussion}\label{sec: Results}
In this section, we discuss the results of the dynamical simulation of both the past and future orbital integrations of the comet 3I's nominal orbit and the statistical clones.

\subsection{Past Evolution}
We evolved the orbit of 3I for a hundred years into the past and found the comet to be at a distance of 1228.58 $\pm$ 0.23 au, originating from the Sagittarius constellation with a geocentric coordinate RA ($\alpha$) and Declination ($\delta$) of 294.9691$^\circ$ $\pm$ 0.0041$^\circ$, and -19.0766$^\circ$ $\pm$ 0.0007$^\circ$ respectively.
The comet is approaching with a radial velocity of -57.995 $\pm$ 0.011 kms$^{-1}$. We have used the \textsc{Astropy} package to convert the equatorial coordinates into Galactic coordinates and Ecliptic coordinates. The Galactic longitude, \textit{l}, is found to be 20.6913$^\circ$ $\pm$ 0.0022$^\circ$ and the Galactic latitude is -18.9947$^\circ$ $\pm$ 0.0033$^\circ$. The Ecliptic longitude ($\lambda$) is 293.530$^\circ$ $\pm$ 0.004$^\circ$, and the Ecliptic latitude ($\beta$) is 2.3459$^\circ$ $\pm$ 0.0001$^\circ$.

The velocities in the ICRS frame, v$_x$, v$_y$ and v$_z$, are found to be -23.198 $\pm$ 0.008, 49.647 $\pm$ 0.008 kms$^{-1}$ and 18.943 $\pm$ 0.003 kms$^{-1}$. The Cartesian velocities are converted into Galactic space velocities using the method given in \citet{Johnson_1987}, where the transformation matrix has been updated at the epoch J2000. The Galactic space velocities ($U$, $V$, $W$) are -51.26 $\pm$ 0.01, -19.40 $\pm$ 0.01, 18.93 $\pm$ 0.01) kms$^{-1}$. Here, we have used the galactocentric sign convention that $U$ is negative in the direction of the Galactic center, $V$ is positive in the direction of Galactic rotation, and $W$ is positive in the direction of the North Galactic Pole (NGP). These velocities have been corrected for the mean solar motion velocities ($U$, $V$, $W$)$_{\odot}$ = ($11.1$, $12.24$, $7.25$) kms$^{-1}$ with respect to the local standard of rest (LSR) \citep{Schonrich_2010}. The corrected velocities of the comet 3I are (62.36 $\pm$ 0.69, -7.16 $\pm$ 0.47, 26.18 $\pm$ 0.36) kms$^{-1}$. These results closely match those reported by \citet{Fuente_3I}.

\citet{Fuente_3I} argue that the comet is originating from the thin disk as the $(U_{LSR}^2+W_{LSR}^2)^{1/2}$ is less than 85 kms$^{-1}$ (based on Fig 3 of \citet{Bensby_2003}).  However, Fig 7 in \citet{Silva2023} shows that the V$_{Total}$ = $(U_{LSR}^2+V_{LSR}^2+W_{LSR}^2)^{1/2}$ range between 50 kms$^{-1}$ and 70 kms$^{-1}$ is characterized as the probable zone of the transition between the thin and thick disk components. For Comet 3I, the corresponding V$_{Ttotal}$ = 68.01 $\pm$ 0.82  kms$^{-1}$. Therefore, we believe that the place of origin of comet 3I is uncertain. This is consistent with the conclusion by \citet{Hopkins_2025}. Further spectroscopic estimation of metallicity for this comet will help to define its place of origin among the thin or thick disk.

\subsection{Future Evolution}
We have divided this subsection into two parts. In the first part, we discuss the long-term future integration of 100 years with the \textsc{Ias15} integrator with a time step of 1 day. In the second part, we have performed a precise simulation of 20 years with a time step of 1 hour using the \textsc{Ias15} integrator.

\subsubsection{Long-term integration}
During the long-term integration, we have simulated the integration of comet 3I and its clones over a period of 100 years. The projected RA ($\alpha$) and Declination ($\delta$) of the comet after 100 years are found to be 95.321$^\circ$ $\pm$ 0.019$^\circ$ and 19.8065$^\circ$ $\pm$ 0.0006$^\circ$, showing that the comet will be leaving towards the Gemini constellation. The radial velocity of the comet will be 58.01 $\pm$ 0.01 kms$^{-1}$. The corresponding Galactic longitude, \textit{l}, will be 191.943$^\circ$ $\pm$ 0.008$^\circ$ and the Galactic latitude will be 2.559$^\circ$ $\pm$ 0.016$^\circ$. The comet 3I will reach a distance of 1221.41 $\pm$ 0.25 au from the Sun at the end of 100 years. The Ecliptic longitude ($\lambda$) and the Ecliptic latitude ($\beta$) will be 95.011$^\circ$ $\pm$ 0.018$^\circ$ and -3.541$^\circ$ $\pm$ 0.001$^\circ$ respectively.

The future velocities in the ICRS frame, v$_x$, v$_y$ and v$_z$, are found to be -4.98 $\pm$ 0.02 kms$^{-1}$, 54.34 $\pm$ 0.01 kms$^{-1}$ and 19.65 $\pm$ 0.01 kms$^{-1}$ which are converted into Galactic space velocities as described in the previous section. The galactic space velocities ($U$, $V$, $W$) are (-56.70 $\pm$ 0.01, -11.95 $\pm$ 0.01, 2.51 $\pm$ 0.02) kms$^{-1}$. The solar motion corrected velocities are (67.80 $\pm$ 0.69, 0.29 $\pm$ 0.47, 9.76 $\pm$ 0.36) kms$^{-1}$.

\subsubsection{Short-term integration}
As we can see in the Figure \ref{fig: orbits}, there are possible close encounters with Mars and Jupiter on epochs JD 2460951.5 (2025 October 03) and JD 2461115.5 (2026 March 16), respectively. The minimum expected distance would be 0.19 au from Mars and 0.357 au from Jupiter. The distance of the comet 3I from Jupiter is very close to the Hill radius (0.355 au) of Jupiter. Therefore, there could be a stronger perturbation from Jupiter compared to Mars.

\begin{figure*}
\gridline{
  \fig{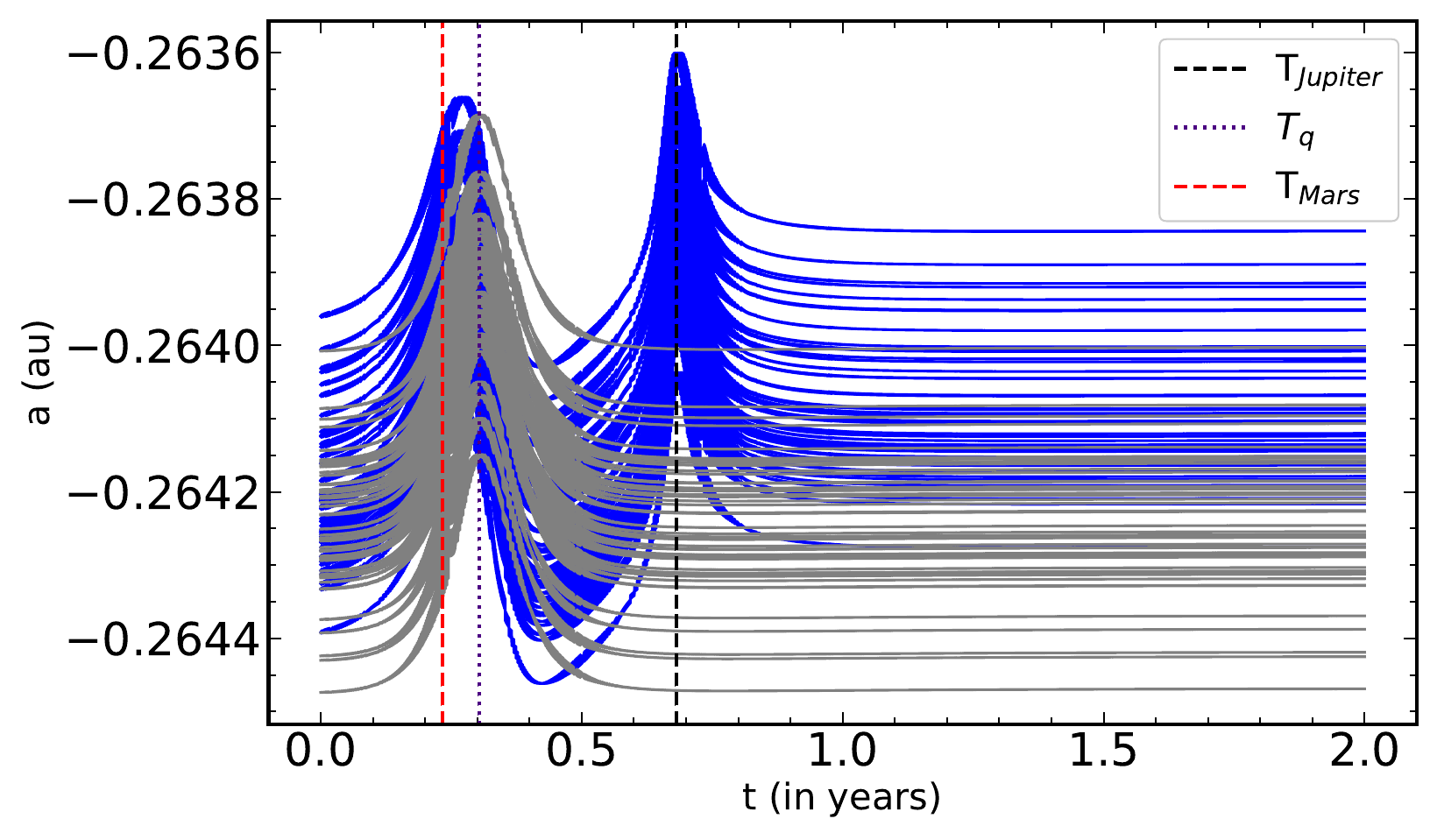}{0.5\textwidth}{(a) Semi-major axis, $a$}
  \fig{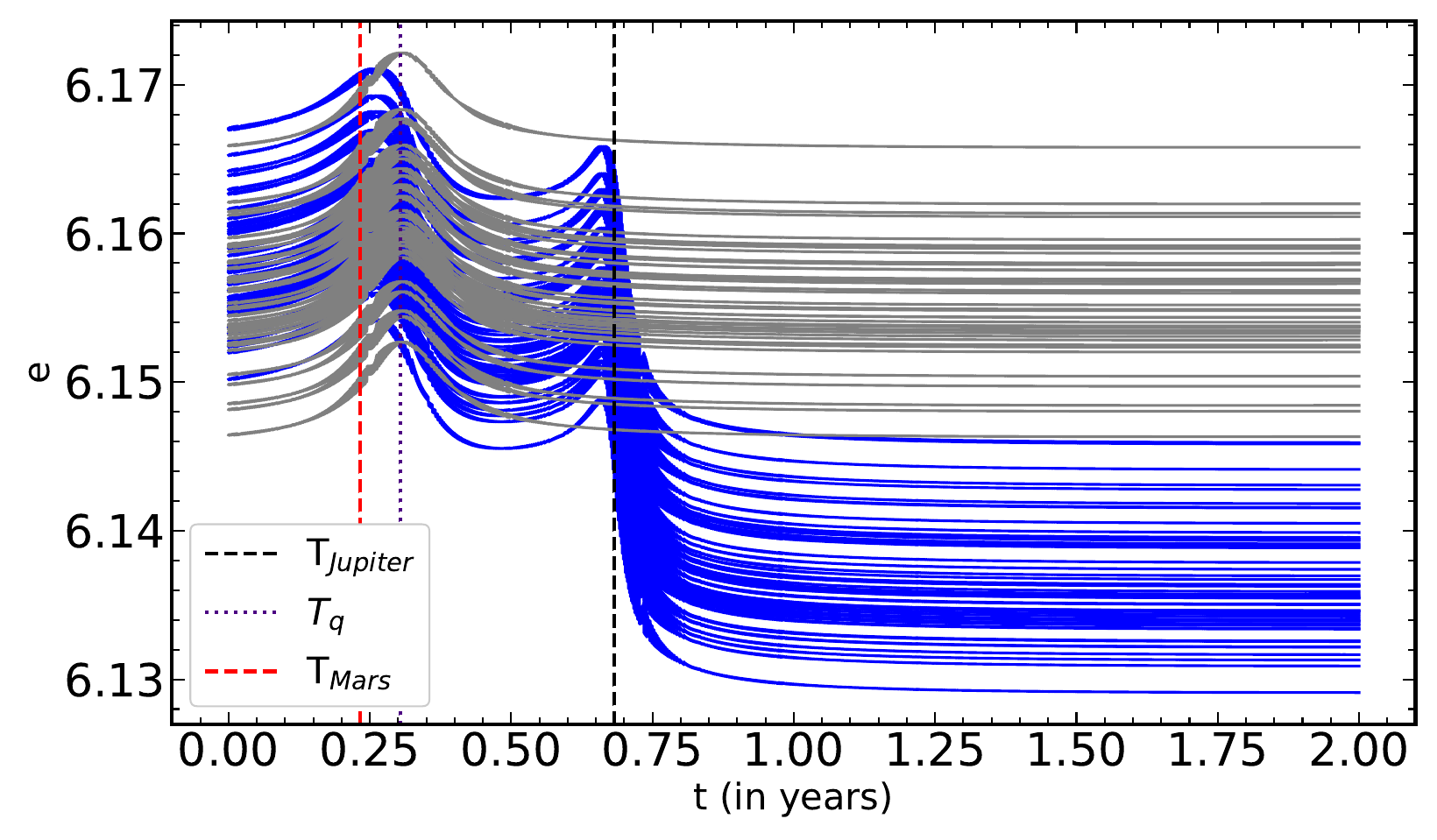}{0.5\textwidth}{(b) Eccentricity, $e$}
}
\gridline{
  \fig{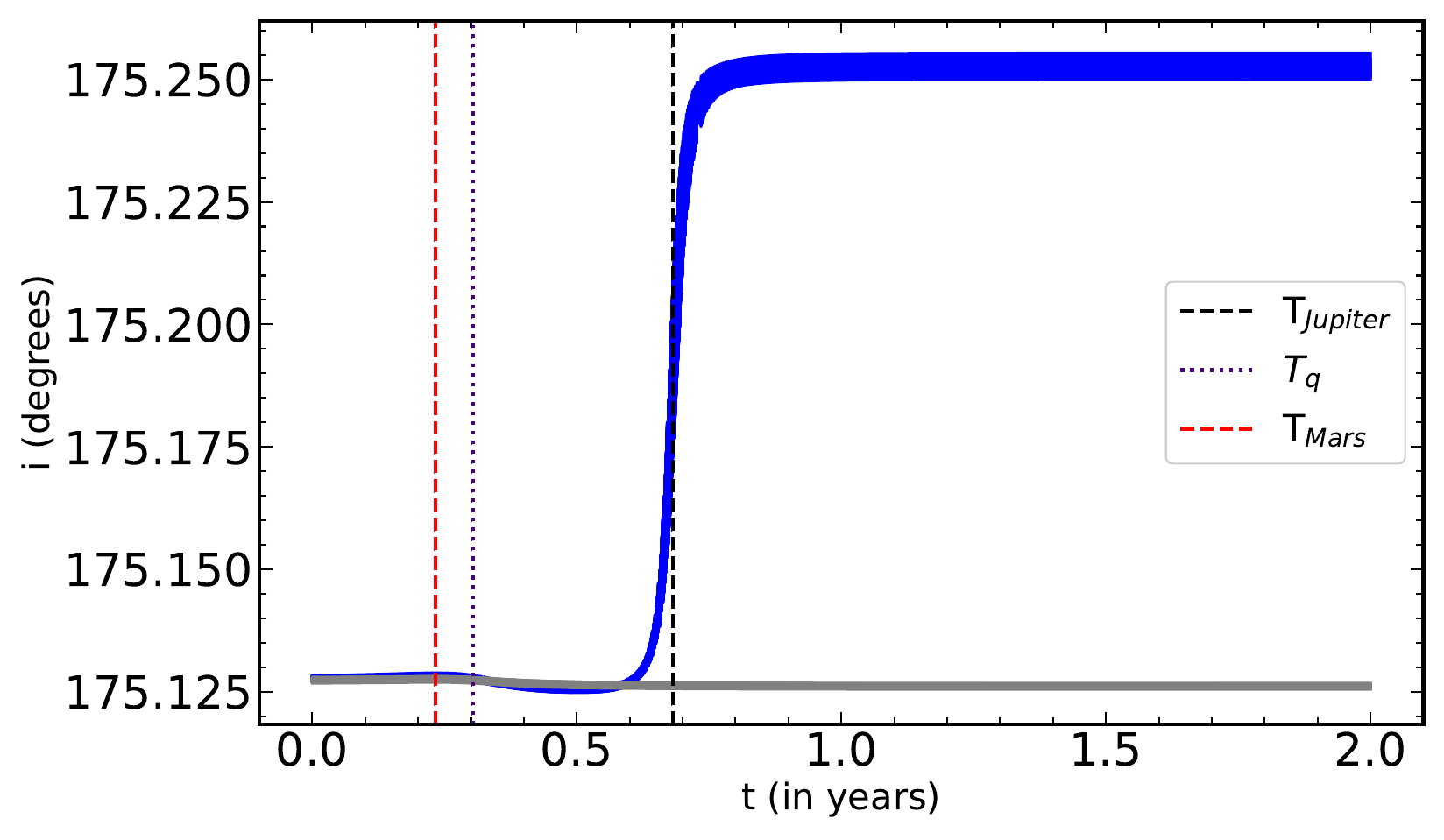}{0.5\textwidth}{(c) Inclination, $i$}
  \fig{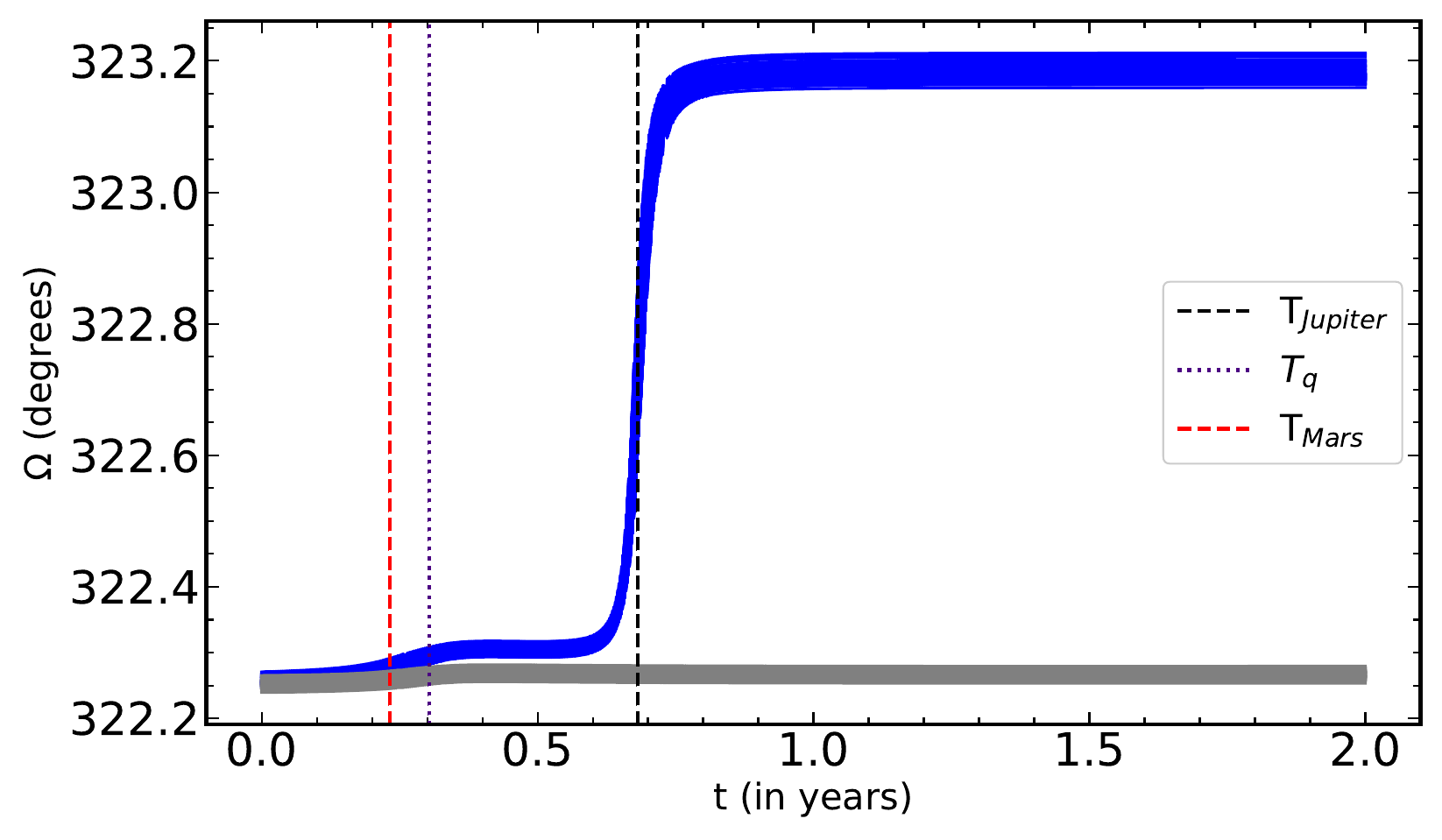}{0.5\textwidth}{(d) Longitude of Ascending Node, $\Omega$}
}
\gridline{
  \fig{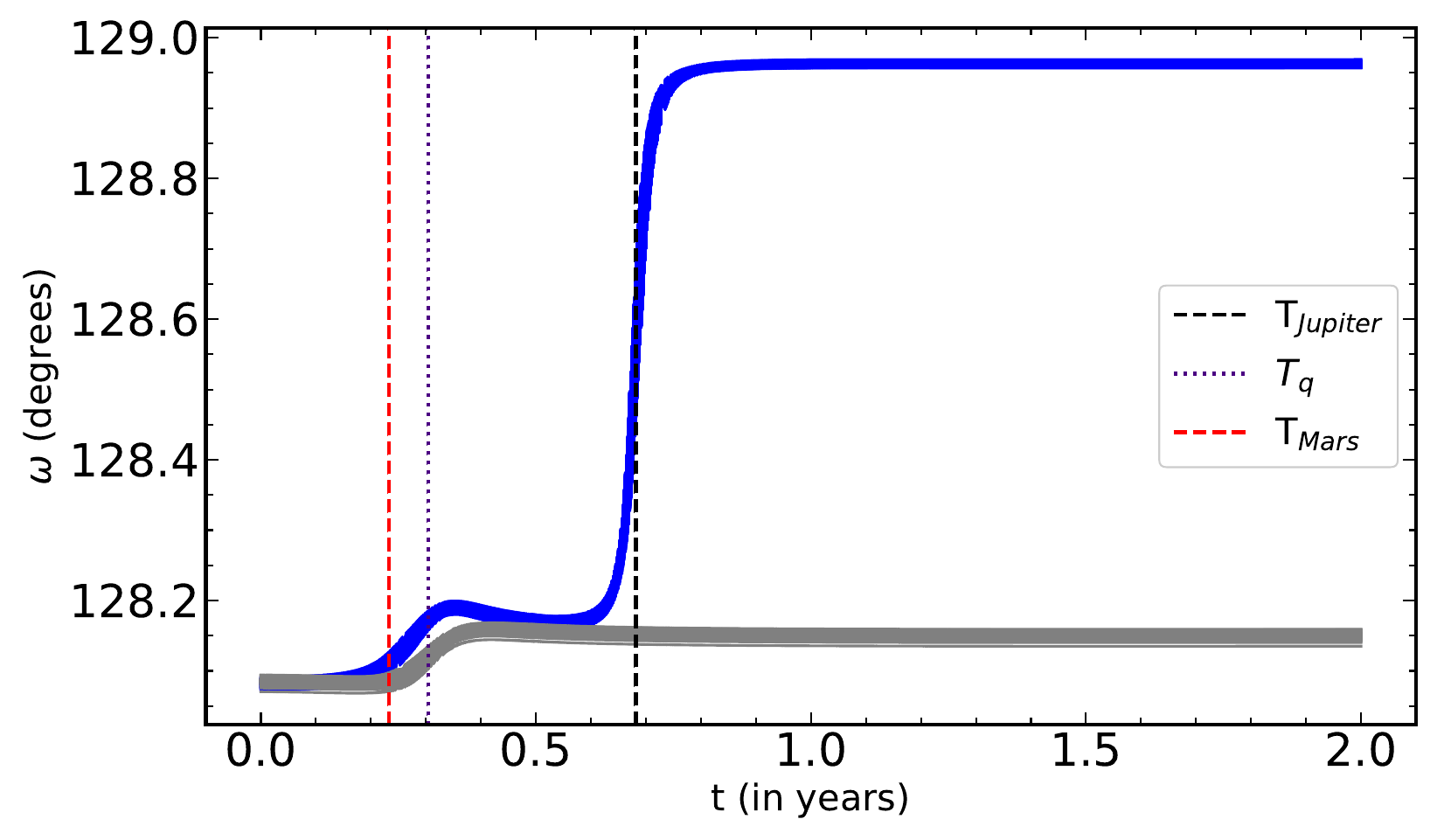}{0.5\textwidth}{(e) Argument of Perihelion, $\omega$}
  \fig{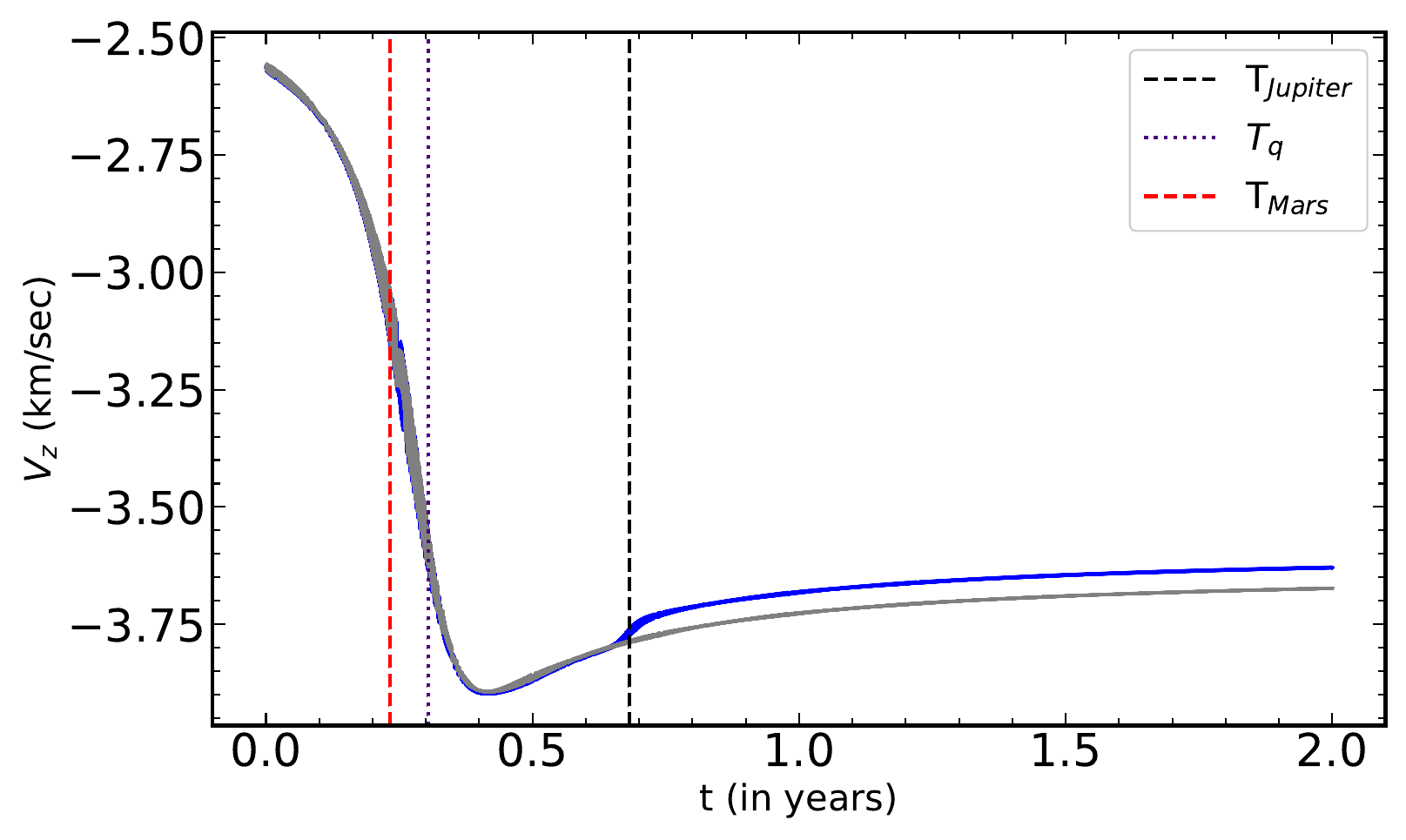}{0.5\textwidth}{(f) Cartesian Velocity in z-direction, $v_z$}
}
\caption{Variation of Orbital Parameters and the Cartesian velocity in z-direction. In each panel, the gray curves are for the simulation without considering interactions with Mars and Jupiter, whereas the blue curves include these planets. The vertical red and black dashed lines represent the date at which the comet is close to Mars and Jupiter. The vertical dotted lines in indigo represent the time of perihelion. Here, we plotted the behavior of 50 clones to keep the image size manageable.}
\label{fig: Orb_Var}
\end{figure*}

In short-term integration, the comet nominal orbit and its statistical clones are integrated for twenty years using the highly precise mathematical integrator \textsc{Ias15} with a time step of 1 hour. 
The variation of the orbital elements such as semi-major axis ($a$), eccentricity ($e$), inclination ($i$), Longitude of Ascending Node ($\Omega$), Argument of Periapsis ($\omega$), and the cartesian velocities in the z-axis are shown in Figure \ref{fig: Orb_Var} for two years. We also checked for post-Newtonian (General Relativistic) correction using the \textsc{gr} force from the REBOUNDX package \citep{reboundx} and didn't see any significant effects affecting the solutions.

\begin{figure}[htbp]
    \centering
    \includegraphics[width=\linewidth]{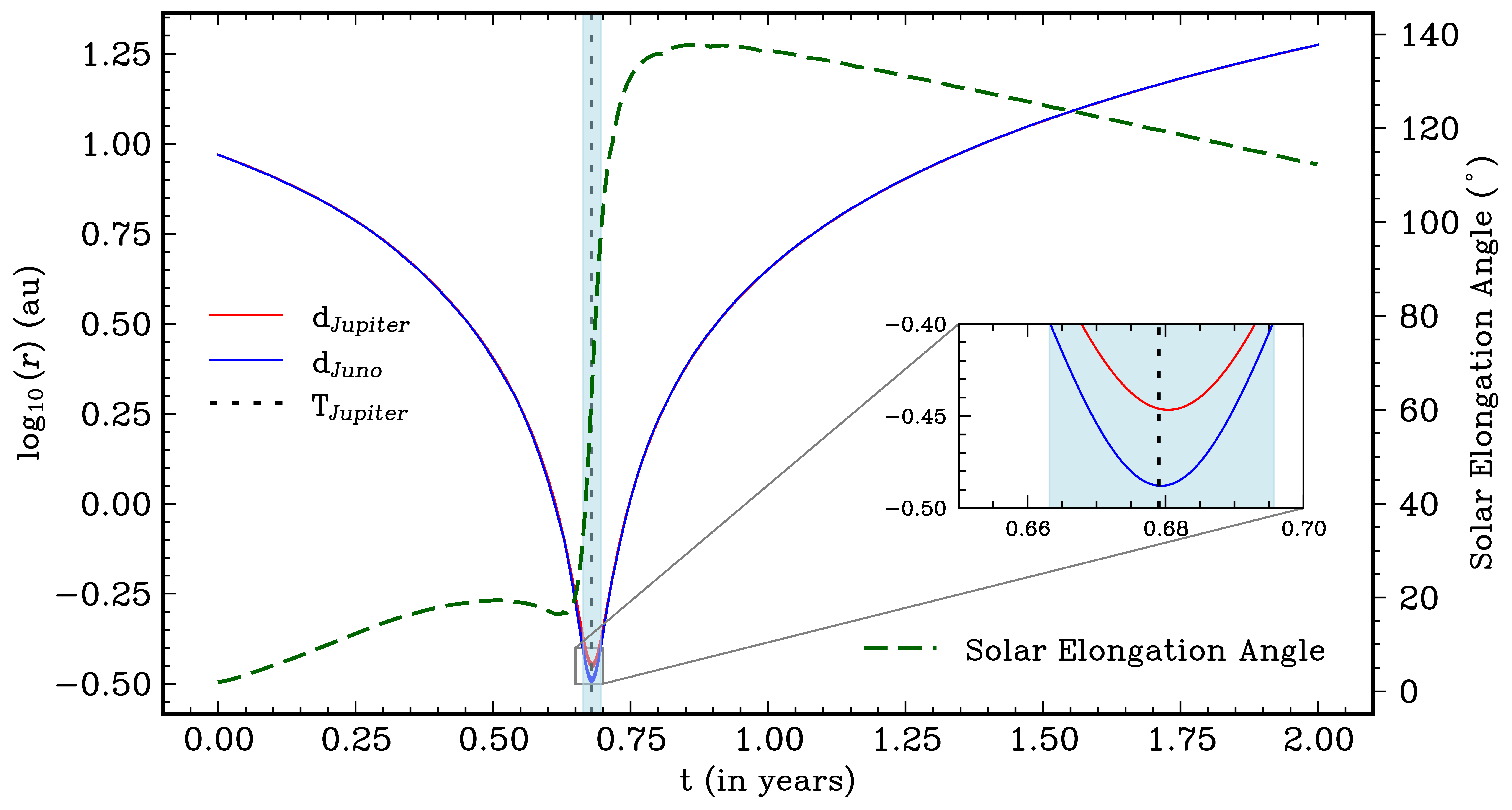}
    \caption{The distance from comet to Juno spacecraft, $d_{Juno}$, and Jupiter $d_{Jupiter}$, are plotted in blue and red curves respectively in log scale. The black dashed line is the time of the interaction with Jupiter. The masked region is for the optimal observation when the comet 3I will be at 0.4 au or below from the Juno spacecraft with a solar elongation between 28$^{\circ}$ to 100$^{\circ}$.}
    \label{fig: Juno_min}
\end{figure}

In the plots, the gray color represents the orbital evolution of cometary clones in the absence of Mars and Jupiter, while the blue color curves represent the effect of both Mars and Jupiter. The vertical red and black dashed line represents the time of close approach with Mars and Jupiter. The vertical, indigo-dotted line represents the time of perihelion. As we can see, there is a strong effect on the orbital parameters, such as semi-major axis ($a$), eccentricity ($e$), inclination ($i$), Longitude of Ascending Node ($\Omega$), and Argument of Periapsis ($\omega$), in the presence of Mars and Jupiter. Also, its z-axis velocity will be significantly perturbed due to the planetary interaction.  The other parameters, namely, cartesian coordinates, i.e., $x$, $y$, $z$, cartesian velocities, $v_x$ and $v_y$, and the sixth orbital element, Mean Anomaly, are not significantly perturbed by the interaction.

Since there is a possibility of close observation from the Juno spacecraft using its suite of onboard instruments, we have considered a different simulation where the Juno spacecraft is added in REBOUND with \textsc{Ias15} integrator. Taking the maximum distance from Juno to be 0.4 au for optimal observation, and the time at which the distance between comet 3I and the Juno spacecraft is less than the distance between comet 3I and Jupiter, we suggest that the optimal period for observation is from 09$^{th}$ to 22$^{nd}$ March 2026. The solar elongation angle (Sun-Juno-Comet angle) during this period will be between 28$^{\circ}$ to 100$^{\circ}$ as shown in Figure \ref{fig: Juno_min}.

\subsubsection{Non-Gravitational Acceleration}\label{sec: NGF}
The comets are subjected to non-gravitational acceleration due to outgassing, as mentioned in the introduction. The non-gravitational forces are defined as:
\begin{equation}
    F_i = A_i g(r) , i = 1,2,3
\end{equation}
where $A_1$ is the non-gravitational acceleration in the radial direction, $A_2$ in the transverse direction, and $A_3$ in the normal direction. $g(r)$ is the semi-empirical function defined as:
\begin{equation}
    g(r) = \alpha \left(\frac{r}{r_0}\right)^{-m} \left[1 + \left(\frac{r}{r_0}\right)^n \right]^{-k}
\end{equation}

Here, different constants such as $\alpha$, $m$, $n$, $k$, $r_0$ are taken from \citet{Królikowska_2017}. These parameters depend on the perihelion distance. If the perihelion distance is less than 4 au, which is true for the comet 3I, the values of constants will be $\alpha = 0.1113$, $m=2.15$, $n=5.093$, $k=4.6142$, $r_0 = 2.808$ au, representing the outgassing due to the water sublimation.  For more details, see \citet{Marsden_1973, sosa_2011, Królikowska_2017}.
Therefore, to predict the effects of non-gravitational forces on comet 3I, we have included the non-gravitational accelerations in all three components in the simulation.  There is another parameter called the DT (time offset in maximum brightness) parameter, representing the asymmetric cometary emission. Since it is difficult to constrain this in advance, we have not included it in our simulation. 

As we write, we have verified that a new solution exists at the epoch JD 2460878.5 (2025-Jul-22.0) TDB; however, the non-gravitational accelerations remain unknown. Hence, we have used the Windows version of the \texttt{Find\_Orb} software, as provided by \citet{Bill_gray_2022}, to fit the orbit and extract the non-gravitational accelerations. We utilize 3,997 observations (May 8 - September 29, 2025) from the Minor Planet Center (MPC) to predict non-gravitational accelerations in different components. We obtain a well-constrained and stable solution to the non-gravitational accelerations only when we consider a symmetric model of comet $A_1$ and $A_2$ parameters with $A_3$ and DT parameters equal to zero.  We find $A_1$ is (1.5 $\pm$ 0.2) $\times$ 10$^{-6}$  and $A_2$ is (2.74 $\pm$ 0.95) $\times$ 10$^{-6}$ respectively, having positive signs for both $A_1$ and $A_2$ components.

We rerun our simulations with different positive values of the non-gravitational accelerations in the $A_1$ and $A_2$ parameters, with values ranging from $10^{-5}$ au day$^ {-2}$ to $10^{-9}$ au day$^ {-2}$.

\begin{figure*}
\gridline{
  \fig{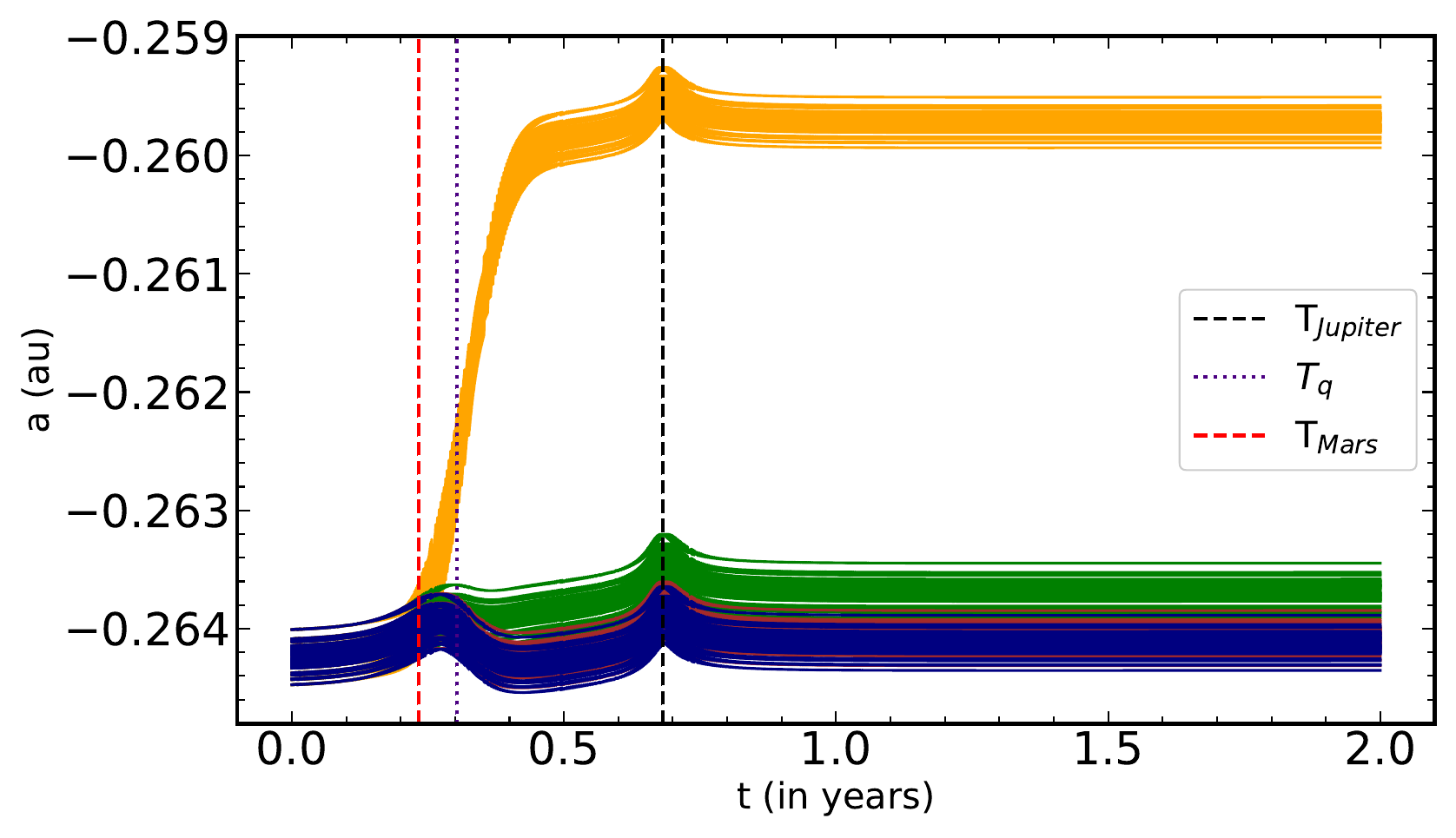}{0.5\textwidth}{(a) Semi-major axis, $a$}
  \fig{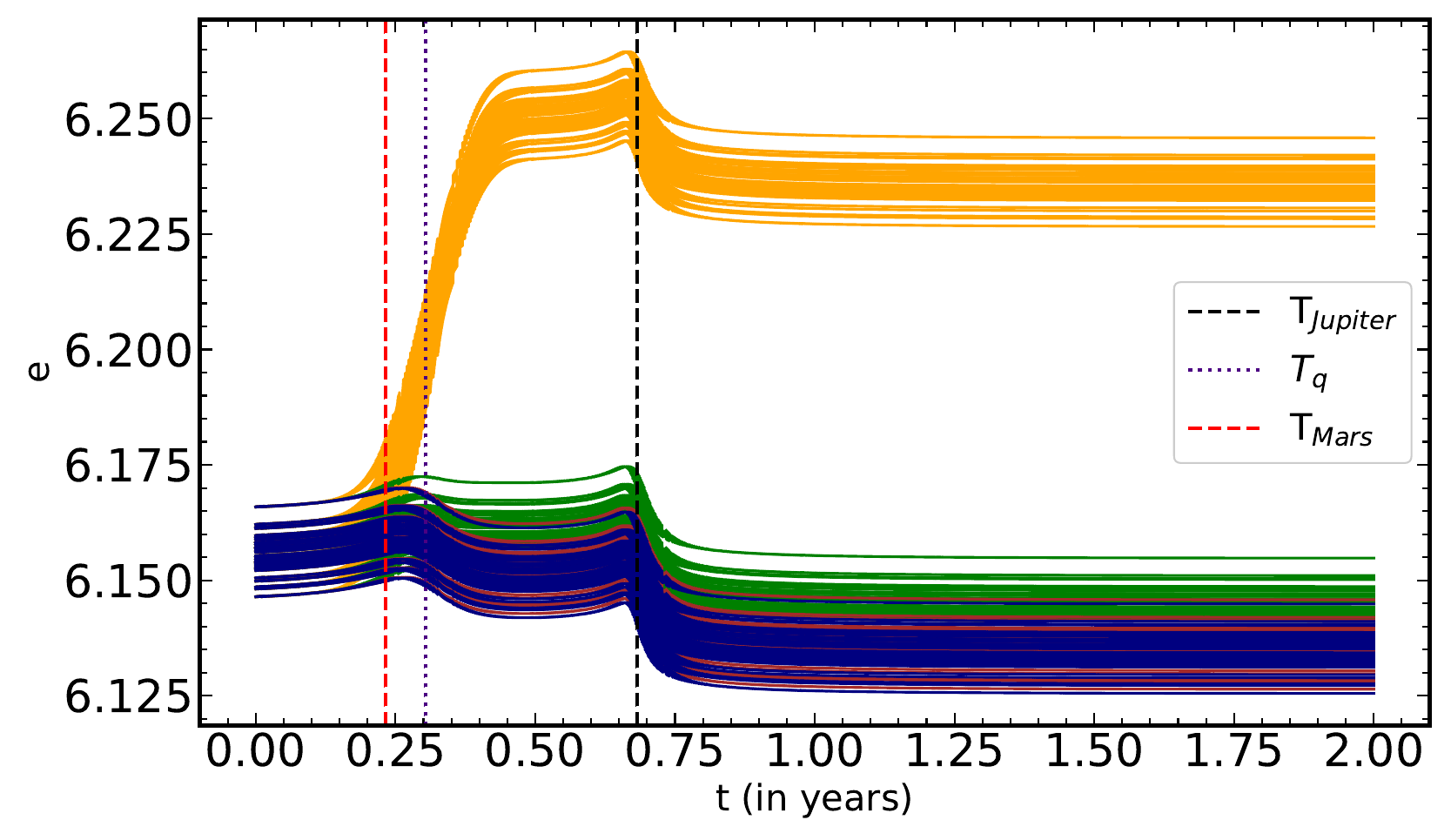}{0.5\textwidth}{(b) Eccentricity, $e$}
}
\gridline{
  \fig{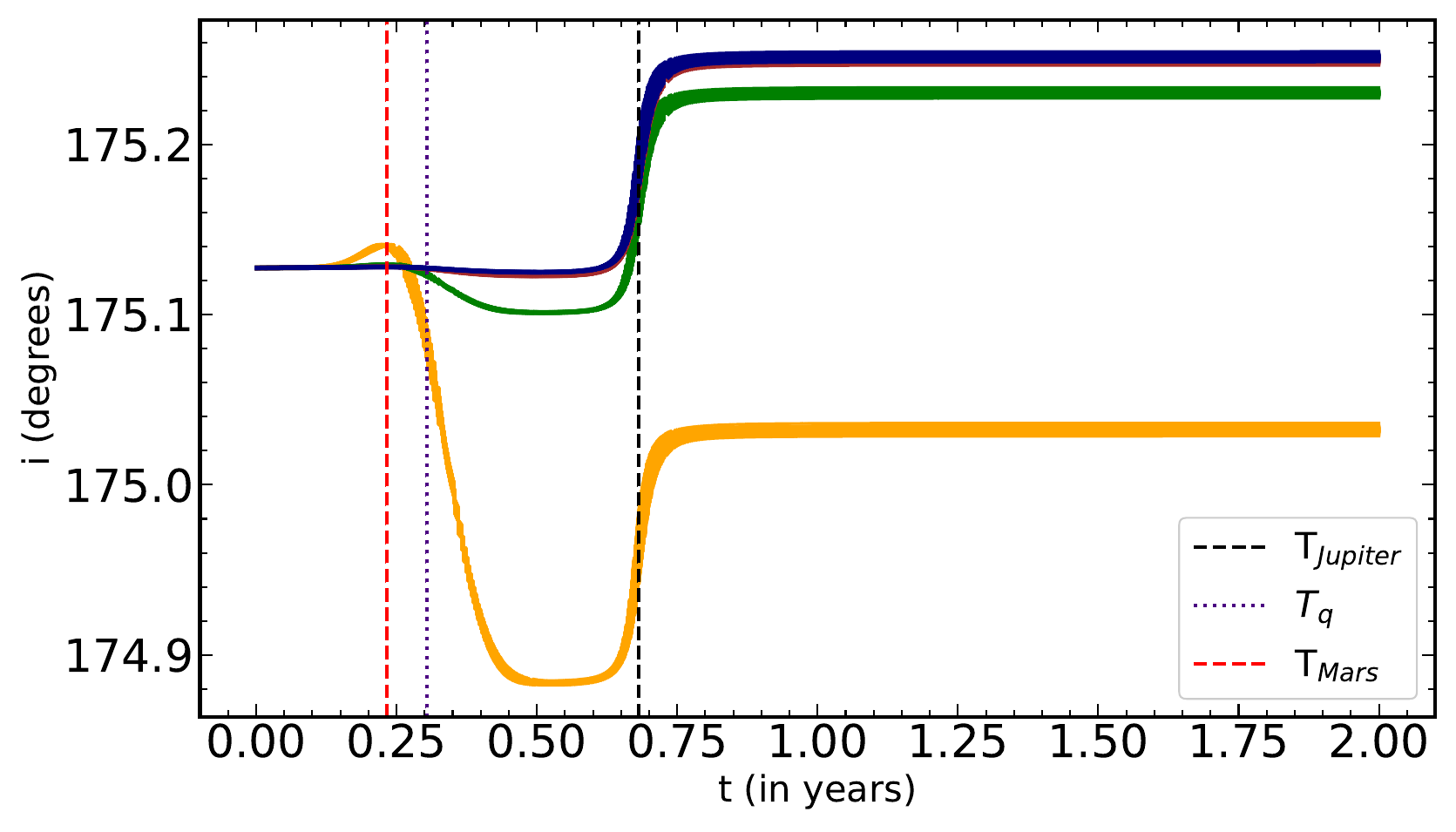}{0.5\textwidth}{(c) Inclination, $i$}
  \fig{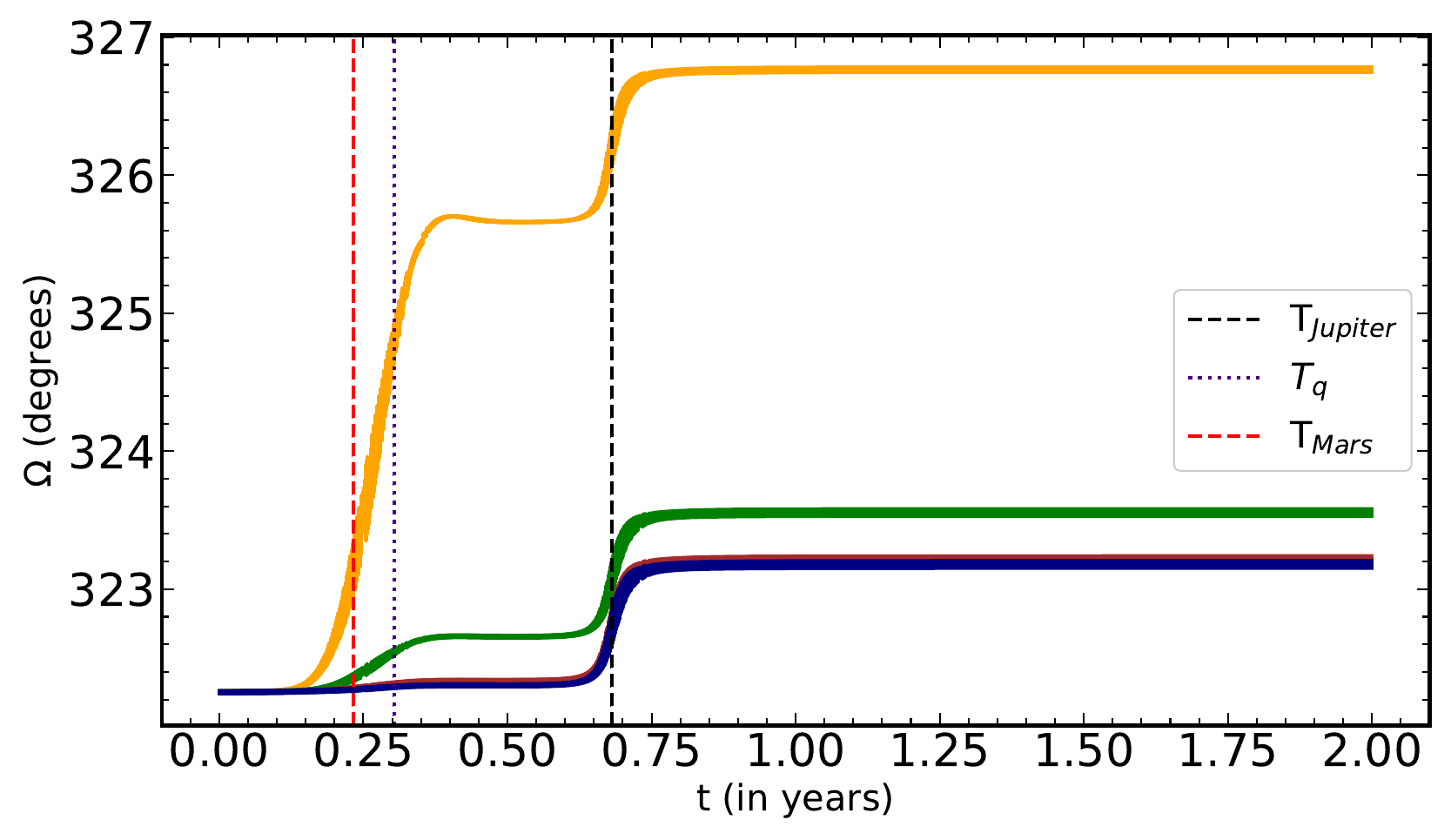}{0.5\textwidth}{(d) Longitude of Ascending Node, $\Omega$}
}
\gridline{
  \fig{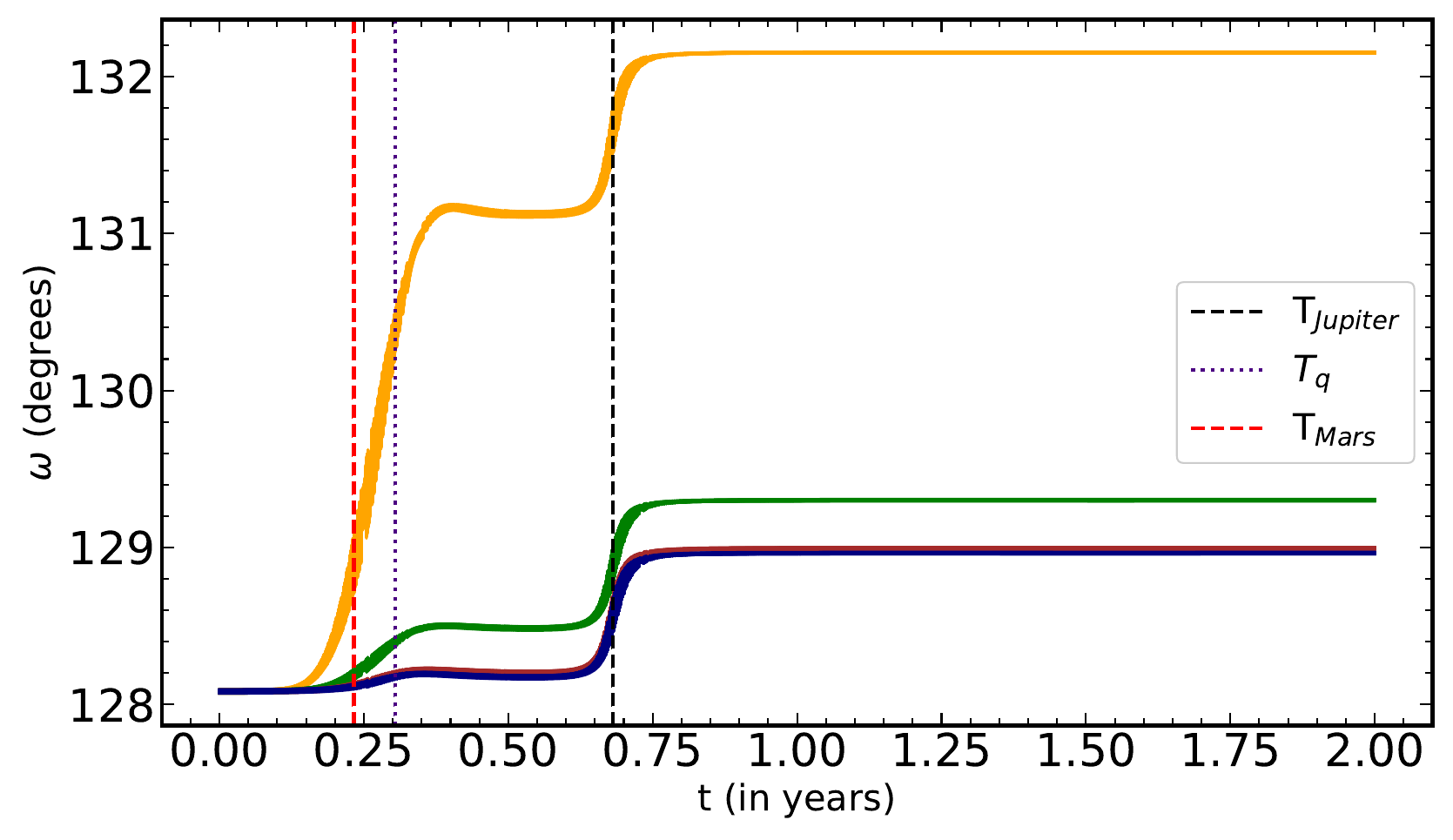}{0.5\textwidth}{(e) Argument of Perihelion, $\omega$}
  \fig{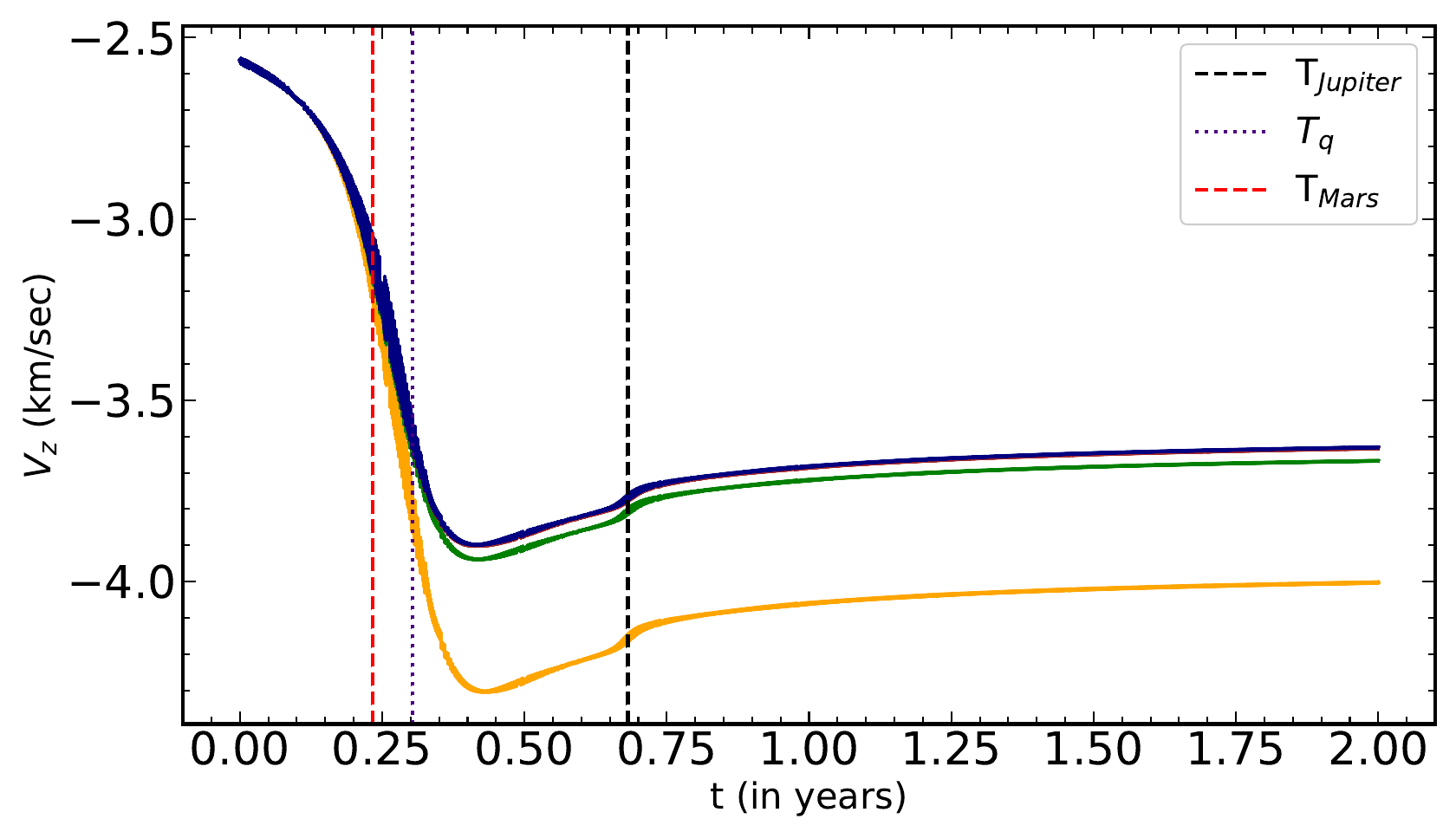}{0.5\textwidth}{(f) Cartesian Velocity in z-direction, $v_z$}
}
\caption{Variation of Orbital Parameters and the Cartesian velocity in z-direction. In each panel, the orange curves represent non - gravitational accelerations of 10$^{-5}$ auday$^{-2}$, the green, brown and navy blue curves represent non - gravitational accelerations of 10$^{-6}$ auday$^{-2}$, 10$^{-7}$ auday$^{-2}$, and 10$^{-8}$ auday$^{-2}$. Here, we plotted the behavior of 50 clones to keep the image size manageable.}
\label{fig: NGF_Orb_Var}
\end{figure*}

We have plotted the results of the variation of the orbital elements semi-major axis ($a$), eccentricity ($e$), inclination ($i$), Longitude of Ascending Node ($\Omega$), Argument of Periapsis ($\omega$), and the cartesian velocity in the z-axis in Figure \ref{fig: NGF_Orb_Var}. As seen in the Figure \ref{fig: NGF_Orb_Var}, the non-gravitational accelerations to show a significant effect on comet 3I are 10$^{-5}$ auday$^{-2}$, 10$^{-6}$ auday$^{-2}$ The non-gravitational acceleration of 10$^{-7}$ auday$^{-2}$ shows a very small effect. The rest of the accelerations below 10$^{-7}$ auday$^{-2}$ show negligible variation from the results given by considering only the gravitational effects. For the strongest non-gravitational acceleration 10$^{-5}$ auday$^{-2}$, the ICRS coordinates:  RA ($\alpha$) and Declination ($\delta$) of the comet after 20 years when the comet reaches close to 250 au, are found to be $96.218^\circ$ $\pm$ $0.015^\circ$ and $19.454^\circ$ $\pm$ $0.0005^\circ$. The Galactic space velocities ($U$, $V$, $W$) are ($-57.09$ $\pm$ $0.01$, $-12.63$ $\pm$ $0.01$, $2.86$ $\pm$ $0.02$) kms$^{-1}$. 

Similarly, for the non-gravitational acceleration of 10$^{-6}$ auday$^{-2}$, the ICRS coordinates:  RA ($\alpha$) and Declination ($\delta$) of the comet are found to be $95.702^\circ$ $\pm$ $0.015^\circ$ and $19.774^\circ$ $\pm$ $0.0005^\circ$. The galactic space velocities ($U$, $V$, $W$) are ($-56.78$ $\pm$ $0.01$, $-12.03$ $\pm$ $0.01$, $2.55$ $\pm$ $0.02$) kms$^{-1}$. For comparison, in the absence of non-gravitational accelerations, the ICRS coordinates:  RA ($\alpha$) and Declination ($\delta$) of the comet after 20 years are found to be $95.634^\circ$ $\pm$ $0.015^\circ$ and $19.8102^\circ$ $\pm$ $0.0005^\circ$. The galactic space velocities ($U$, $V$, $W$) are ($-56.74$ $\pm$ $0.01$, $-11.96$ $\pm$ $0.01$, $2.51$ $\pm$ $0.01$) kms$^{-1}$.

We have also examined the variation in the Cartesian coordinates $x$, and $z$, as well as the Cartesian velocities $v_x$ and $v_y$. There is no visible effect in the $y$ coordinate. The major effect of the non-gravitational acceleration is seen for the values of 10$^{-5}$ auday$^{-2}$, while the non-gravitational accelerations below 10$^{-5}$ auday$^{-2}$ have no significant effect (See Figure \ref{fig: NGF_Orb_Var_1} in the appendix section \ref{ap_NGF}). 

\subsubsection{Comparing the distribution of clones with or without non-gravitational accelerations.}
\begin{figure}[htbp]
\centering
\begin{subfigure}[b]{0.49\textwidth}
\centering
\includegraphics[width=\textwidth]{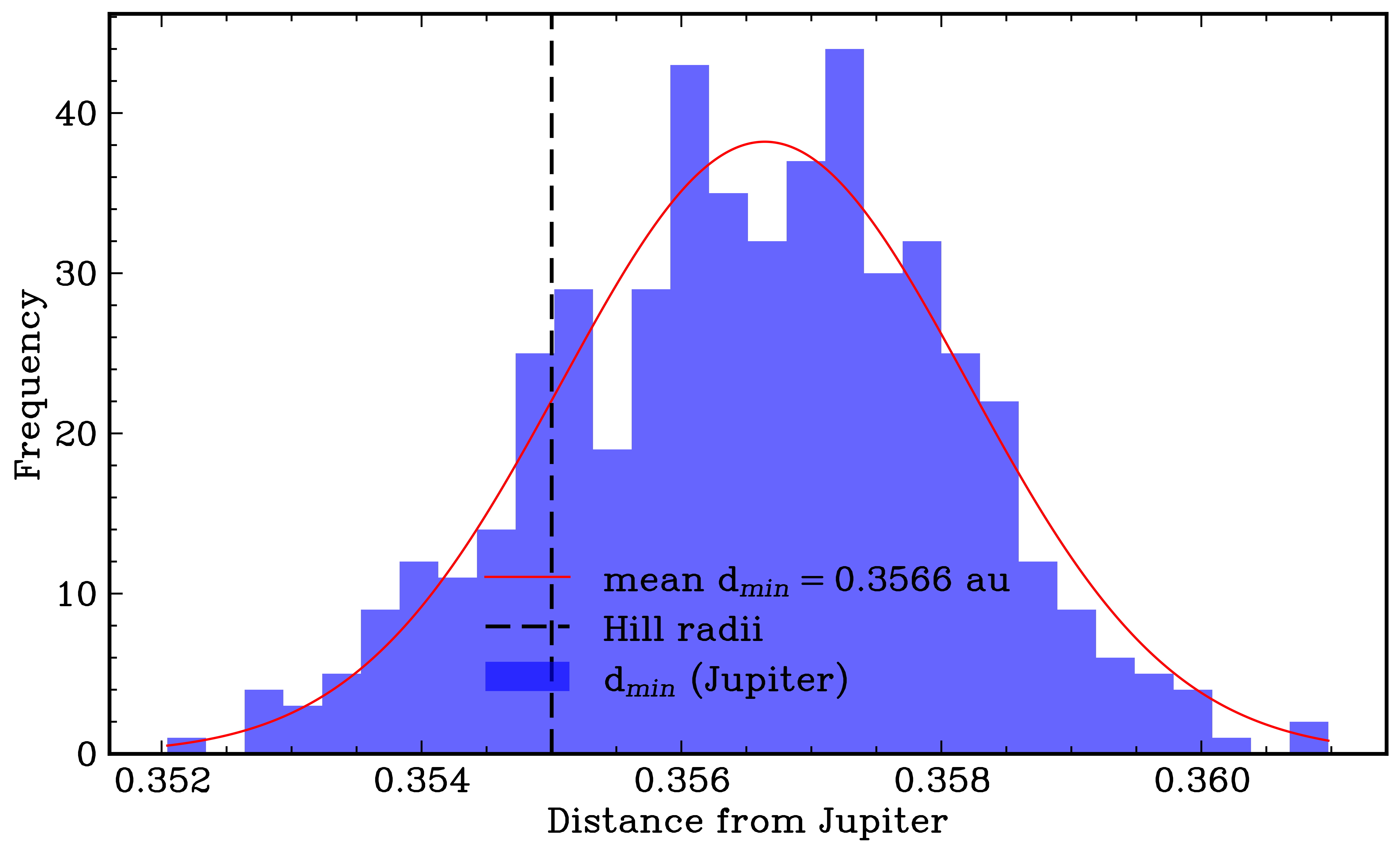}
\caption{Histogram of the minimum distance from Jupiter for simulations with 500 clones in the presence of gravitational forces only.}
\label{fig: histogram_GR}
\end{subfigure}
\hfill
\begin{subfigure}[b]{0.49\textwidth}
\centering
\includegraphics[width=\textwidth]{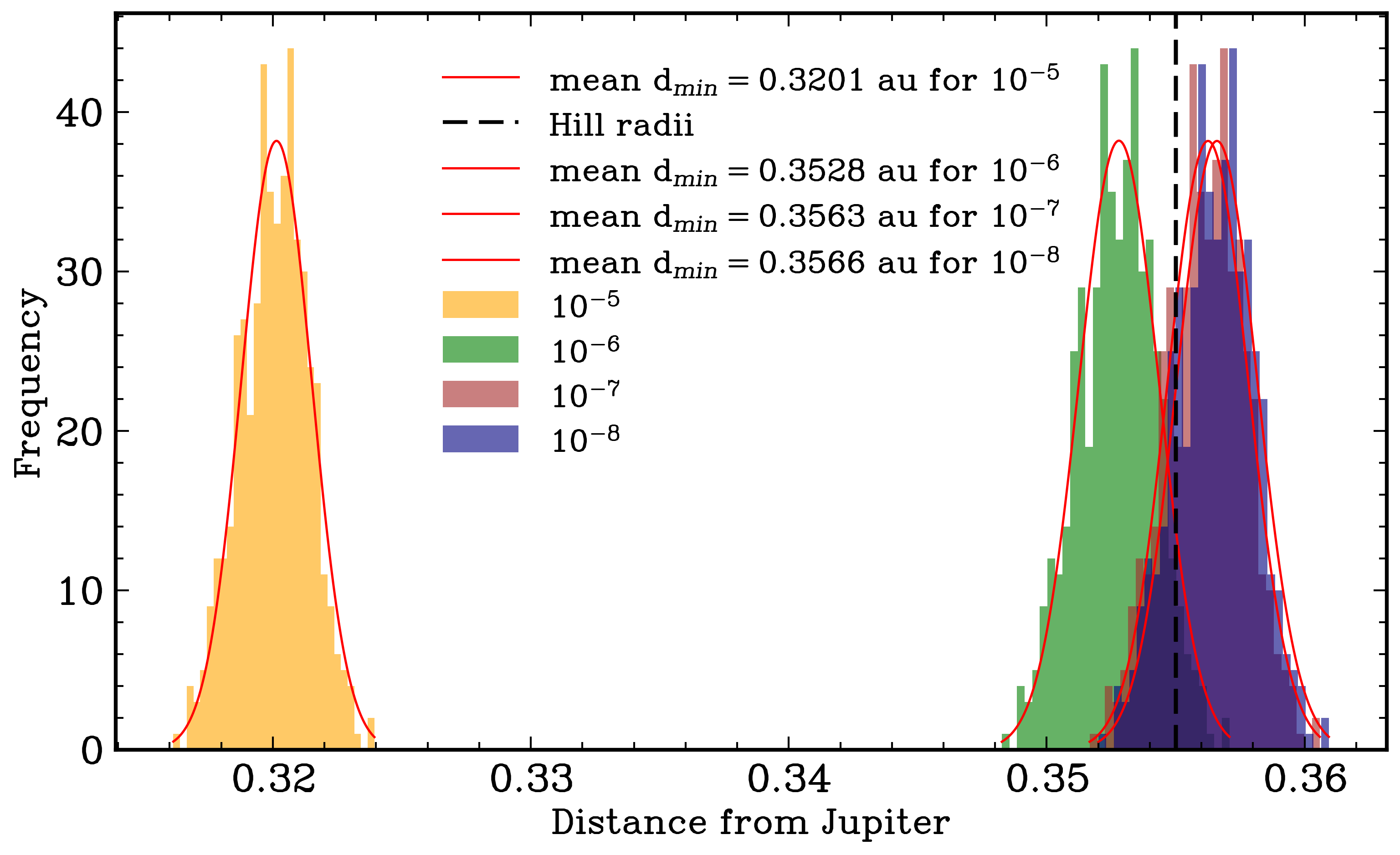}
\caption{Histogram of the minimum distance from Jupiter for simulations with 500 clones in the presence of both gravitational and non-gravitational forces. The different non-gravitational accelerations are marked in the same colours as the previous plots.}
\label{fig: histogram_NG}
\end{subfigure}

\caption{Histogram of 500 cometary clones in the presence of different forces.}
\label{fig:main}

\end{figure}

Since there could be a strong effect from Jupiter, we plotted the histogram of the distribution of the close approach distances of the cometary clones when they approach Jupiter. In the first case, we consider the effect of gravitational forces only. As shown in Figure \ref{fig: histogram_GR}, the mean minimum distance for clones from Jupiter is 0.3566 au with a 1-$\sigma$ of 0.0016 au. We find that 81 out of 500 clones pass at distances lesser than the Hill radius of Jupiter.

For different non-gravitational forces, the mean of the close-approach distance changes significantly, when we consider the non-gravitational acceleration of 10$^{-5}$ auday$^{-2}$, the mean value of the minimum distance is $0.3201$ and 100 $\%$ of clones are insider the Hill radius of Jupiter while for non-gravitational acceleration of 10$^{-6}$ auday$^{-2}$ 470 out of 500 clones are inside the Hill radius of Jupiter with a mean value of $0.3528$ (see Figure \ref{fig: histogram_NG}). For non-gravitational acceleration of 10$^{-7}$ auday$^{-2}$, 116 out of 500 clones are inside the Hill radius of Jupiter with a mean value of $0.3563$, and for 10$^{-8}$ auday$^{-2}$, 87 out of 500 clones are inside the Hill radius of Jupiter with a mean value of $0.3566$.

\section{Summary and Conclusion}\label{sec: Conclusion}
In this letter, we have studied the effects of both gravitational perturbations from Mars and Jupiter, as well as non-gravitational forces resulting from cometary emission, on comet 3I. We have found the following:

\begin{itemize}
    \item The comet 3I will definitely suffer the perturbation from both Mars and Jupiter at their respective close approach epochs. The effect of Jupiter will be larger due to the fact that the comet is passing very close to the Hill radius of Jupiter.
    \item The long-term orbital integration of 500 statistical clones of comet 3I for hundred years in the past and future shows that the comet is coming from the Sagittarius constellation with a mean radial velocity of \text-57.995 $\pm$ 0.011 kms$^{-1}$  and leaving towards the Gemini constellation with a mean radial velocity of 58.01 $\pm$ 0.01 kms$^{-1}$.
    \item The solar motion corrected Galactic space velocities $(U, V, W)$ at 1 hundred years ago are (62.36 $\pm$ 0.69, -7.16 $\pm$ 0.47, 26.18 $\pm$ 0.36) kms$^{-1}$. Correspondingly, the comet 3I's 
    $(U_{LSR}^2+V_{LSR}^2+W_{LSR}^2)^{1/2}$ is 68.01 $\pm$ 0.82  kms$^{-1}$. From the Toomre diagram mentioned in the \citet{Silva2023}, we found that the comet total velocity, $V_{Total}$, $(U_{LSR}^2+V_{LSR}^2+W_{LSR}^2)^{1/2}$, is between 50 to 70 kms$^{-1}$. Hence, the possible location of the comet 3I is found to be in the transition zone from the thin to the thick disk. Therefore, it is uncertain to confirm the region of origin using kinematic information.
    \item The short-term orbital integration for 20 years in the presence of gravitational perturbation affects the various orbital parameters and Cartesian velocity in the z-axis significantly, as shown in Figure \ref{fig: Orb_Var}.
    \item A stable solution for the non-gravitational accelerations using \texttt{Find\_Orb} has been derived when we consider the symmetric model. The non-gravitational accelerations, $A_1$ and $A_2$, are found to be (1.5 $\pm$ 0.2) $\times$ 10$^{-6}$  and (2.74 $\pm$ 0.95) $\times$ 10$^{-6}$ respectively.
    \item The integration in the presence of non-gravitational forces can significantly affect the orbital parameters, resulting in a change in the future orbital evolution of the comet 3I, as shown in Figures \ref{fig: NGF_Orb_Var} and \ref{fig: NGF_Orb_Var_1}. We find that the major effects on comet 3I's orbit are seen when the non-gravitational acceleration is of the order of 10$^{-5}$ to 10$^{-6}$ auday$^{-2}$. The non-gravitational acceleration of 10$^{-7}$ auday$^{-2}$ and lower has negligible effect on the various parameters.
    \item The optimal window of the observation from the Juno spacecraft will be from 09$^{th}$ to 22$^{nd}$ March 2026. During this time, the solar elongation angle will be between 28$^{\circ}$ to 100$^{\circ}$.
    \item Further observations of the comet 3I are required to confirm the non-gravitational accelerations. Additionally, if the light curve of comet 3I is asymmetric, then the DT (time offset in the maximum brightness) parameter will be crucial in calculating the effect of non-gravitational forces.
    
\end{itemize}
\begin{acknowledgments}
We thank the anonymous reviewer for providing valuable comments and helpful suggestions. GA would like to thank Prof. Raul de la Fuente Marcos for the discussion on the Galactocentric coordinate system. We would also like to thank Prof. Agnes Fienga for the discussion on the numerical simulation and useful suggestions. Work at the Physical Research Laboratory is supported by the Department of Space, Govt. of India. The computations were performed on the Param Vikram-1000 High Performance Computing Cluster of the Physical Research Laboratory (PRL).
\end{acknowledgments}

\begin{contribution}

GA: Conceptualization, Methodology, Software, Formal analysis, Visualization, Writing $-$ original draft.

SG: Supervision, Writing $-$ review \& editing.


\end{contribution}

%


\software{Astropy \citep{2013A&A...558A..33A,2018AJ....156..123A,2022ApJ...935..167A},  
          Astroquery \citep{astroquery_2019}, 
          Numpy \citep{Numpy_2020},
          Scipy \citep{2020SciPy-NMeth},
          Matplotlib \citep{Matplotlib_2007},
          Smplotlib \citep{smplotlib},
          Pandas \citep{pandas_2020},
          REBOUND \citep{rebound,reboundias15},
          REBOUNDX \citep{reboundx},
          Find\_Orb \citep{Bill_gray_2022}
          }

\appendix

\section{Variation of different parameters considering non-gravitational solution}\label{ap_NGF}
Here, we show the variation of the Cartesian coordinates, i.e., $x$, and $z$, and the Cartesian velocities in the x-direction, $v_x$, and the y-direction, $v_y$, of 50 statistical clones (for representation purposes only) for two years under the influence of both gravitational and non-gravitational forces.
\begin{figure*}
\gridline{
  \fig{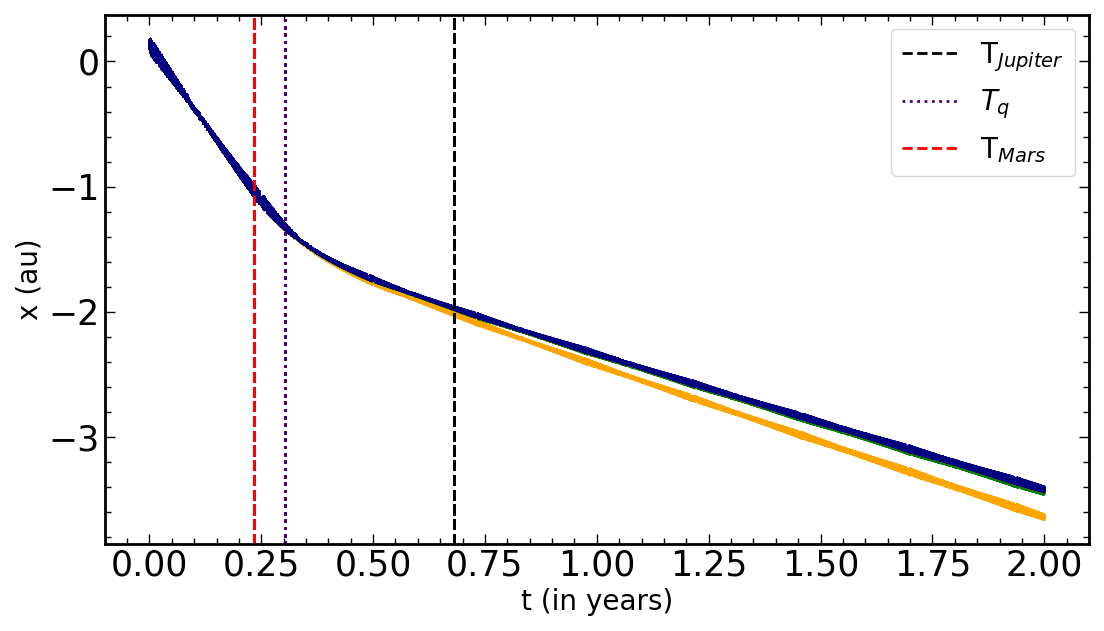}{0.5\textwidth}{(a) Cartesian Coordinate in x-direction, $x$}
  \fig{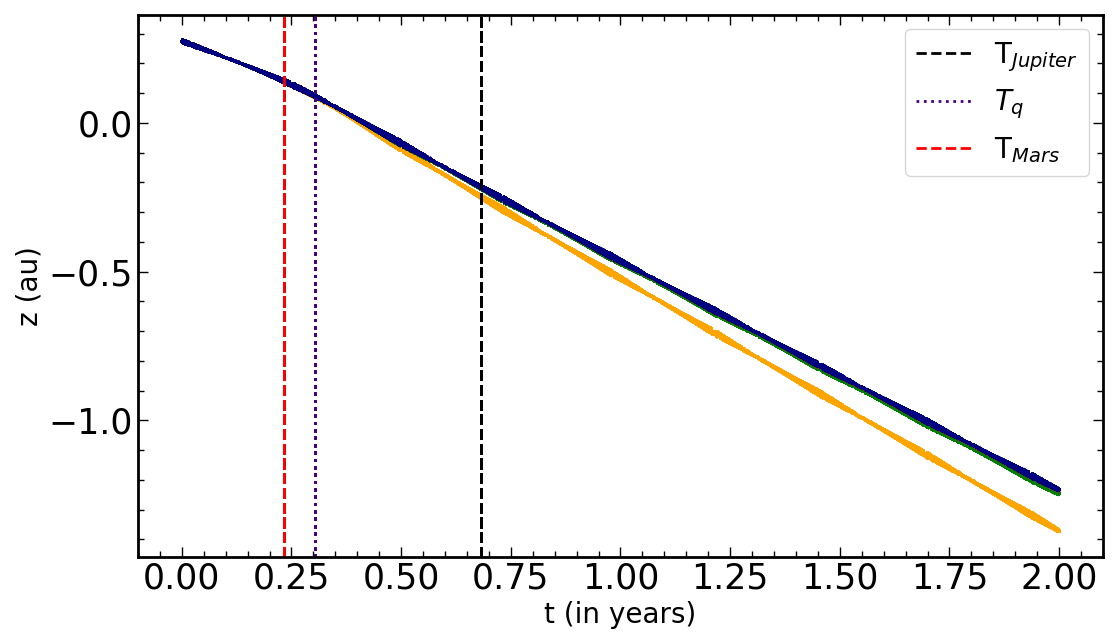}{0.5\textwidth}{(c) Cartesian Coordinate in z-direction, $z$}
}
\gridline{
  \fig{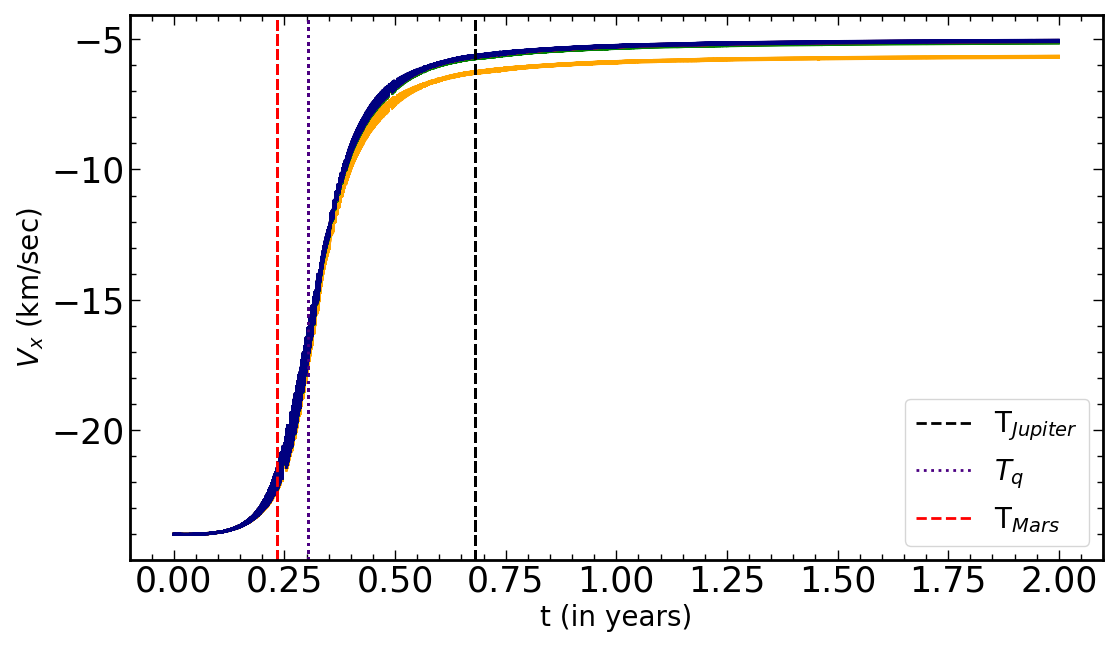}{0.5\textwidth}{(d) Cartesian Velocity in x-direction, $v_x$}
  \fig{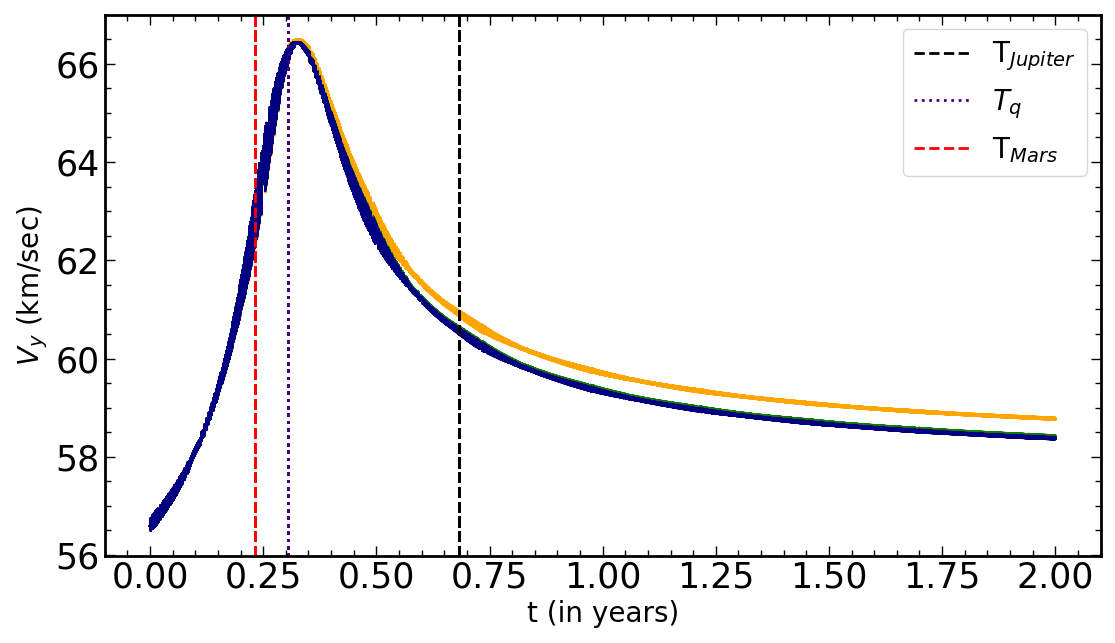}{0.5\textwidth}{(e) Cartesian Velocity in y-direction, $v_y$}
}
\caption{Variation of Cartesian Coordinates, $x$, and $z$, the Cartesian velocities, $v_x$, and $v_y$. In each panel, the orange curves represent non - gravitational accelerations of 10$^{-5}$ auday$^{-2}$, the green, brown and navy blue curves represent non - gravitational accelerations of 10$^{-6}$ auday$^{-2}$, 10$^{-7}$ auday$^{-2}$, and 10$^{-8}$ auday$^{-2}$.  Here, we plotted the behavior of 50 clones to keep the image size manageable.}
\label{fig: NGF_Orb_Var_1}
\end{figure*}


\bibliography{Reference}{}

\begin{thebibliography}{}
\expandafter\ifx\csname natexlab\endcsname\relax\def\natexlab#1{#1}\fi
\providecommand{\url}[1]{\href{#1}{#1}}
\providecommand{\dodoi}[1]{doi:~\href{http://doi.org/#1}{\nolinkurl{#1}}}
\providecommand{\doeprint}[1]{\href{http://ascl.net/#1}{\nolinkurl{http://ascl.net/#1}}}
\providecommand{\doarXiv}[1]{\href{https://arxiv.org/abs/#1}{\nolinkurl{https://arxiv.org/abs/#1}}}

\bibitem[{S. {Anderson} {et~al.}(2024){Anderson}, {Rousselot}, {Jehin}, {Noyelles}, {Manfroid}, {Hardy}, \& {Robert}}]{Anderson_2024}
{Anderson}, S., {Rousselot}, P., {Jehin}, E., {et~al.} 2024, \bibinfo{title}{{Comet C/1908 R1 (Morehouse) as a C/2016 R2 (PanSTARRS)-like comet},} in European Planetary Science Congress, EPSC2024--59, \dodoi{10.5194/epsc2024-59}

\bibitem[{S.~E. Anderson {et~al.}(2023)Anderson, Rousselot, Noyelles, Jehin, \& Mousis}]{sarah_multi_2023}
Anderson, S.~E., Rousselot, P., Noyelles, B., Jehin, E., \& Mousis, O. 2023, \bibinfo{title}{N2/CO ratio in comets insensitive to orbital evolution,} Monthly Notices of the Royal Astronomical Society, 524, 5182, \dodoi{10.1093/mnras/stad2092}

\bibitem[{K. Aravind {et~al.}(2021)Aravind, Ganesh, Venkataramani, Sahu, Angchuk, Sivarani, \& Unni}]{Aravind_2I_2021}
Aravind, K., Ganesh, S., Venkataramani, K., {et~al.} 2021, \bibinfo{title}{Activity of the first interstellar comet 2I/Borisov around perihelion: results from Indian observatories,} Monthly Notices of the Royal Astronomical Society, 502, 3491, \dodoi{10.1093/mnras/stab084}

\bibitem[{ {Astropy Collaboration} {et~al.}(2013){Astropy Collaboration}, {Robitaille}, {Tollerud}, {Greenfield}, {Droettboom}, {Bray}, {Aldcroft}, {Davis}, {Ginsburg}, {Price-Whelan}, {Kerzendorf}, {Conley}, {Crighton}, {Barbary}, {Muna}, {Ferguson}, {Grollier}, {Parikh}, {Nair}, {Unther}, {Deil}, {Woillez}, {Conseil}, {Kramer}, {Turner}, {Singer}, {Fox}, {Weaver}, {Zabalza}, {Edwards}, {Azalee Bostroem}, {Burke}, {Casey}, {Crawford}, {Dencheva}, {Ely}, {Jenness}, {Labrie}, {Lim}, {Pierfederici}, {Pontzen}, {Ptak}, {Refsdal}, {Servillat}, \& {Streicher}}]{2013A&A...558A..33A}
{Astropy Collaboration}, {Robitaille}, T.~P., {Tollerud}, E.~J., {et~al.} 2013, \bibinfo{title}{{Astropy: A community Python package for astronomy},} \aap, 558, A33, \dodoi{10.1051/0004-6361/201322068}

\bibitem[{ {Astropy Collaboration} {et~al.}(2018){Astropy Collaboration}, {Price-Whelan}, {Sip{\H{o}}cz}, {G{\"u}nther}, {Lim}, {Crawford}, {Conseil}, {Shupe}, {Craig}, {Dencheva}, {Ginsburg}, {VanderPlas}, {Bradley}, {P{\'e}rez-Su{\'a}rez}, {de Val-Borro}, {Aldcroft}, {Cruz}, {Robitaille}, {Tollerud}, {Ardelean}, {Babej}, {Bach}, {Bachetti}, {Bakanov}, {Bamford}, {Barentsen}, {Barmby}, {Baumbach}, {Berry}, {Biscani}, {Boquien}, {Bostroem}, {Bouma}, {Brammer}, {Bray}, {Breytenbach}, {Buddelmeijer}, {Burke}, {Calderone}, {Cano Rodr{\'\i}guez}, {Cara}, {Cardoso}, {Cheedella}, {Copin}, {Corrales}, {Crichton}, {D'Avella}, {Deil}, {Depagne}, {Dietrich}, {Donath}, {Droettboom}, {Earl}, {Erben}, {Fabbro}, {Ferreira}, {Finethy}, {Fox}, {Garrison}, {Gibbons}, {Goldstein}, {Gommers}, {Greco}, {Greenfield}, {Groener}, {Grollier}, {Hagen}, {Hirst}, {Homeier}, {Horton}, {Hosseinzadeh}, {Hu}, {Hunkeler}, {Ivezi{\'c}}, {Jain}, {Jenness}, {Kanarek}, {Kendrew}, {Kern}, {Kerzendorf}, {Khvalko}, {King}, {Kirkby}, {Kulkarni},
  {Kumar}, {Lee}, {Lenz}, {Littlefair}, {Ma}, {Macleod}, {Mastropietro}, {McCully}, {Montagnac}, {Morris}, {Mueller}, {Mumford}, {Muna}, {Murphy}, {Nelson}, {Nguyen}, {Ninan}, {N{\"o}the}, {Ogaz}, {Oh}, {Parejko}, {Parley}, {Pascual}, {Patil}, {Patil}, {Plunkett}, {Prochaska}, {Rastogi}, {Reddy Janga}, {Sabater}, {Sakurikar}, {Seifert}, {Sherbert}, {Sherwood-Taylor}, {Shih}, {Sick}, {Silbiger}, {Singanamalla}, {Singer}, {Sladen}, {Sooley}, {Sornarajah}, {Streicher}, {Teuben}, {Thomas}, {Tremblay}, {Turner}, {Terr{\'o}n}, {van Kerkwijk}, {de la Vega}, {Watkins}, {Weaver}, {Whitmore}, {Woillez}, {Zabalza}, \& {Astropy Contributors}}]{2018AJ....156..123A}
{Astropy Collaboration}, {Price-Whelan}, A.~M., {Sip{\H{o}}cz}, B.~M., {et~al.} 2018, \bibinfo{title}{{The Astropy Project: Building an Open-science Project and Status of the v2.0 Core Package},} \aj, 156, 123, \dodoi{10.3847/1538-3881/aabc4f}

\bibitem[{ {Astropy Collaboration} {et~al.}(2022){Astropy Collaboration}, {Price-Whelan}, {Lim}, {Earl}, {Starkman}, {Bradley}, {Shupe}, {Patil}, {Corrales}, {Brasseur}, {N{\"o}the}, {Donath}, {Tollerud}, {Morris}, {Ginsburg}, {Vaher}, {Weaver}, {Tocknell}, {Jamieson}, {van Kerkwijk}, {Robitaille}, {Merry}, {Bachetti}, {G{\"u}nther}, {Aldcroft}, {Alvarado-Montes}, {Archibald}, {B{\'o}di}, {Bapat}, {Barentsen}, {Baz{\'a}n}, {Biswas}, {Boquien}, {Burke}, {Cara}, {Cara}, {Conroy}, {Conseil}, {Craig}, {Cross}, {Cruz}, {D'Eugenio}, {Dencheva}, {Devillepoix}, {Dietrich}, {Eigenbrot}, {Erben}, {Ferreira}, {Foreman-Mackey}, {Fox}, {Freij}, {Garg}, {Geda}, {Glattly}, {Gondhalekar}, {Gordon}, {Grant}, {Greenfield}, {Groener}, {Guest}, {Gurovich}, {Handberg}, {Hart}, {Hatfield-Dodds}, {Homeier}, {Hosseinzadeh}, {Jenness}, {Jones}, {Joseph}, {Kalmbach}, {Karamehmetoglu}, {Ka{\l}uszy{\'n}ski}, {Kelley}, {Kern}, {Kerzendorf}, {Koch}, {Kulumani}, {Lee}, {Ly}, {Ma}, {MacBride}, {Maljaars}, {Muna}, {Murphy}, {Norman},
  {O'Steen}, {Oman}, {Pacifici}, {Pascual}, {Pascual-Granado}, {Patil}, {Perren}, {Pickering}, {Rastogi}, {Roulston}, {Ryan}, {Rykoff}, {Sabater}, {Sakurikar}, {Salgado}, {Sanghi}, {Saunders}, {Savchenko}, {Schwardt}, {Seifert-Eckert}, {Shih}, {Jain}, {Shukla}, {Sick}, {Simpson}, {Singanamalla}, {Singer}, {Singhal}, {Sinha}, {Sip{\H{o}}cz}, {Spitler}, {Stansby}, {Streicher}, {{\v{S}}umak}, {Swinbank}, {Taranu}, {Tewary}, {Tremblay}, {de Val-Borro}, {Van Kooten}, {Vasovi{\'c}}, {Verma}, {de Miranda Cardoso}, {Williams}, {Wilson}, {Winkel}, {Wood-Vasey}, {Xue}, {Yoachim}, {Zhang}, {Zonca}, \& {Astropy Project Contributors}}]{2022ApJ...935..167A}
{Astropy Collaboration}, {Price-Whelan}, A.~M., {Lim}, P.~L., {et~al.} 2022, \bibinfo{title}{{The Astropy Project: Sustaining and Growing a Community-oriented Open-source Project and the Latest Major Release (v5.0) of the Core Package},} \apj, 935, 167, \dodoi{10.3847/1538-4357/ac7c74}

\bibitem[{T. {Bensby} {et~al.}(2003){Bensby}, {Feltzing}, \& {Lundstr{\"o}m}}]{Bensby_2003}
{Bensby}, T., {Feltzing}, S., \& {Lundstr{\"o}m}, I. 2003, \bibinfo{title}{{Elemental abundance trends in the Galactic thin and thick disks as traced by nearby F and G dwarf stars},} \aap, 410, 527, \dodoi{10.1051/0004-6361:20031213}

\bibitem[{D. {Bodewits} {et~al.}(2020){Bodewits}, {Noonan}, {Feldman}, {Bannister}, {Farnocchia}, {Harris}, {Li}, {Mandt}, {Parker}, \& {Xing}}]{Bodewits_2I_2020}
{Bodewits}, D., {Noonan}, J.~W., {Feldman}, P.~D., {et~al.} 2020, \bibinfo{title}{{The carbon monoxide-rich interstellar comet 2I/Borisov},} Nature Astronomy, 4, 867, \dodoi{10.1038/s41550-020-1095-2}

\bibitem[{B.~T. Bolin {et~al.}(2025)Bolin, Belyakov, Fremling, Graham, Elhosseiny, Gray, Ingebretsen, Jewett, Lisse, Karpov, Kilic, Mašek, Molham, Roderick, Takey, Abron, Coughlin, Hsieh, Noll, \& Wong}]{3I_disc_1}
Bolin, B.~T., Belyakov, M., Fremling, C., {et~al.} 2025, \bibinfo{title}{Interstellar comet 3I/ATLAS: discovery and physical description,} Monthly Notices of the Royal Astronomical Society: Letters, slaf078, \dodoi{10.1093/mnrasl/slaf078}

\bibitem[{M.~P. da Silva {et~al.}(2023)da Silva, Alves-Brito, \& do Nascimento}]{Silva2023}
da Silva, M.~P., Alves-Brito, A., \& do Nascimento, J~D, J. 2023, \bibinfo{title}{Stellar rotations and the disentanglement of the Galactic thin and thick discs,} Monthly Notices of the Royal Astronomical Society, 527, 11082, \dodoi{10.1093/mnras/stad3937}

\bibitem[{C. de~la Fuente~Marcos \& R. de~la Fuente~Marcos(2012)de~la Fuente~Marcos \& de~la Fuente~Marcos}]{Fuente_2012_2002VE68}
de~la Fuente~Marcos, C., \& de~la Fuente~Marcos, R. 2012, \bibinfo{title}{On the dynamical evolution of 2002 VE68,} Monthly Notices of the Royal Astronomical Society, 427, 728, \dodoi{10.1111/j.1365-2966.2012.21936.x}

\bibitem[{C. de~la Fuente~Marcos \& R. de~la Fuente~Marcos(2014)de~la Fuente~Marcos \& de~la Fuente~Marcos}]{Fuente_2014_ND15}
de~la Fuente~Marcos, C., \& de~la Fuente~Marcos, R. 2014, \bibinfo{title}{Asteroid 2013 ND15: Trojan companion to Venus, PHA to the Earth,} Monthly Notices of the Royal Astronomical Society, 439, 2970, \dodoi{10.1093/mnras/stu152}

\bibitem[{R. {de la Fuente Marcos} {et~al.}(2025){de la Fuente Marcos}, {Alarcon}, {Licandro}, {Serra-Ricart}, {de Le{\'o}n}, {de la Fuente Marcos}, {Lombardi}, {Tejero}, {Cabrera-Lavers}, {Guerra Arencibia}, \& {Ruiz Cejudo}}]{Fuente_3I}
{de la Fuente Marcos}, R., {Alarcon}, M.~R., {Licandro}, J., {et~al.} 2025, \bibinfo{title}{{Assessing interstellar comet 3I/ATLAS with the 10.4 m Gran Telescopio Canarias and the Two-meter Twin Telescope},} \aap, 700, L9, \dodoi{10.1051/0004-6361/202556439}

\bibitem[{P.~A. Dybczyński \& M. Królikowska(2011)Dybczyński \& Królikowska}]{Dynczynski_2011_250au}
Dybczyński, P.~A., \& Królikowska, M. 2011, \bibinfo{title}{Where do long-period comets come from? Moving through the Jupiter–Saturn barrier,} Monthly Notices of the Royal Astronomical Society, 416, 51, \dodoi{10.1111/j.1365-2966.2011.19005.x}

\bibitem[{A. Egal {et~al.}(2022)Egal, Wiegert, \& Brown}]{Egal_multivar_2022}
Egal, A., Wiegert, P., \& Brown, P.~G. 2022, \bibinfo{title}{A proposed alternative dynamical history for 2P/Encke that explains the taurid meteoroid complex,} Monthly Notices of the Royal Astronomical Society, 515, 2800, \dodoi{10.1093/mnras/stac1839}

\bibitem[{A. {Fitzsimmons} {et~al.}(2019){Fitzsimmons}, {Hainaut}, {Meech}, {Jehin}, {Moulane}, {Opitom}, {Yang}, {Keane}, {Kleyna}, {Micheli}, \& {Snodgrass}}]{Fitzsimmons_2I_2019}
{Fitzsimmons}, A., {Hainaut}, O., {Meech}, K.~J., {et~al.} 2019, \bibinfo{title}{{Detection of CN Gas in Interstellar Object 2I/Borisov},} \apjl, 885, L9, \dodoi{10.3847/2041-8213/ab49fc}

\bibitem[{R. {Gabryszewski} {et~al.}(2024){Gabryszewski}, {Wajer}, \& {W{\l}odarczyk}}]{Gab_2024}
{Gabryszewski}, R., {Wajer}, P., \& {W{\l}odarczyk}, I. 2024, \bibinfo{title}{{Main-belt comets as contributors to the near-Earth objects population},} \aap, 691, A130, \dodoi{10.1051/0004-6361/202347278}

\bibitem[{A. {Ginsburg} {et~al.}(2019){Ginsburg}, {Sip{\H{o}}cz}, {Brasseur}, {Cowperthwaite}, {Craig}, {Deil}, {Guillochon}, {Guzman}, {Liedtke}, {Lian Lim}, {Lockhart}, {Mommert}, {Morris}, {Norman}, {Parikh}, {Persson}, {Robitaille}, {Segovia}, {Singer}, {Tollerud}, {de Val-Borro}, {Valtchanov}, {Woillez}, {Astroquery Collaboration}, \& {a subset of astropy Collaboration}}]{astroquery_2019}
{Ginsburg}, A., {Sip{\H{o}}cz}, B.~M., {Brasseur}, C.~E., {et~al.} 2019, \bibinfo{title}{{astroquery: An Astronomical Web-querying Package in Python},} \aj, 157, 98, \dodoi{10.3847/1538-3881/aafc33}

\bibitem[{J.~D. {Giorgini} {et~al.}(1996){Giorgini}, {Yeomans}, {Chamberlin}, {Chodas}, {Jacobson}, {Keesey}, {Lieske}, {Ostro}, {Standish}, \& {Wimberly}}]{JPL}
{Giorgini}, J.~D., {Yeomans}, D.~K., {Chamberlin}, A.~B., {et~al.} 1996, \bibinfo{title}{{JPL's On-Line Solar System Data Service},} in AAS/Division for Planetary Sciences Meeting Abstracts, Vol.~28, AAS/Division for Planetary Sciences Meeting Abstracts \#28, 25.04

\bibitem[{B. {Gray}(2022){Gray}}]{Bill_gray_2022}
{Gray}, B. 2022, {Find\_Orb: Orbit determination from observations},, Astrophysics Source Code Library, record ascl:2202.016 \doeprint{2202.016}

\bibitem[{P. {Guzik} {et~al.}(2020){Guzik}, {Drahus}, {Rusek}, {Waniak}, {Cannizzaro}, \& {Pastor-Marazuela}}]{Piotr_2I_Nature2020}
{Guzik}, P., {Drahus}, M., {Rusek}, K., {et~al.} 2020, \bibinfo{title}{{Initial characterization of interstellar comet 2I/Borisov},} Nature Astronomy, 4, 53, \dodoi{10.1038/s41550-019-0931-8}

\bibitem[{C.~R. Harris {et~al.}(2020)Harris, Millman, van~der Walt, Gommers, Virtanen, Cournapeau, Wieser, Taylor, Berg, Smith, Kern, Picus, Hoyer, van Kerkwijk, Brett, Haldane, del R{\'{i}}o, Wiebe, Peterson, G{\'{e}}rard-Marchant, Sheppard, Reddy, Weckesser, Abbasi, Gohlke, \& Oliphant}]{Numpy_2020}
Harris, C.~R., Millman, K.~J., van~der Walt, S.~J., {et~al.} 2020, \bibinfo{title}{Array programming with {NumPy},} Nature, 585, 357, \dodoi{10.1038/s41586-020-2649-2}

\bibitem[{M.~J. {Hopkins} {et~al.}(2025){Hopkins}, {Dorsey}, {Forbes}, {Bannister}, {Lintott}, \& {Leicester}}]{Hopkins_2025}
{Hopkins}, M.~J., {Dorsey}, R.~C., {Forbes}, J.~C., {et~al.} 2025, \bibinfo{title}{{From a Different Star: 3I/ATLAS in the Context of the {\={O}}tautahi{\textendash}Oxford Interstellar Object Population Model},} \apjl, 990, L30, \dodoi{10.3847/2041-8213/adfbf4}

\bibitem[{M.-T. Hui \& D. Jewitt(2017)Hui \& Jewitt}]{Hui_2017}
Hui, M.-T., \& Jewitt, D. 2017, \bibinfo{title}{NON-GRAVITATIONAL ACCELERATION OF THE ACTIVE ASTEROIDS,} The Astronomical Journal, 153, 80, \dodoi{10.3847/1538-3881/153/2/80}

\bibitem[{J.~D. Hunter(2007)Hunter}]{Matplotlib_2007}
Hunter, J.~D. 2007, \bibinfo{title}{Matplotlib: A 2D graphics environment,} Computing in Science \& Engineering, 9, 90, \dodoi{10.1109/MCSE.2007.55}

\bibitem[{D. Jewitt {et~al.}(2020)Jewitt, Hui, Kim, Mutchler, Weaver, \& Agarwal}]{Jewitt_2I_2020}
Jewitt, D., Hui, M.-T., Kim, Y., {et~al.} 2020, \bibinfo{title}{The Nucleus of Interstellar Comet 2I/Borisov,} The Astrophysical Journal Letters, 888, L23, \dodoi{10.3847/2041-8213/ab621b}

\bibitem[{D. {Jewitt} \& D.~Z. {Seligman}(2023){Jewitt} \& {Seligman}}]{2023ARA&A..61..197J}
{Jewitt}, D., \& {Seligman}, D.~Z. 2023, \bibinfo{title}{{The Interstellar Interlopers},} \araa, 61, 197, \dodoi{10.1146/annurev-astro-071221-054221}

\bibitem[{D.~R.~H. {Johnson} \& D.~R. {Soderblom}(1987){Johnson} \& {Soderblom}}]{Johnson_1987}
{Johnson}, D. R.~H., \& {Soderblom}, D.~R. 1987, \bibinfo{title}{{Calculating Galactic Space Velocities and Their Uncertainties, with an Application to the Ursa Major Group},} \aj, 93, 864, \dodoi{10.1086/114370}

\bibitem[{M. Królikowska \& P.~A. Dybczyński(2010)Królikowska \& Dybczyński}]{Królikowska_2010_250au}
Królikowska, M., \& Dybczyński, P.~A. 2010, \bibinfo{title}{Where do long-period comets come from? 26 comets from the non-gravitational Oort spike,} Monthly Notices of the Royal Astronomical Society, 404, 1886, \dodoi{10.1111/j.1365-2966.2010.16403.x}

\bibitem[{M. Królikowska \& P.~A. Dybczyński(2017)Królikowska \& Dybczyński}]{Królikowska_2017}
Królikowska, M., \& Dybczyński, P.~A. 2017, \bibinfo{title}{Oort spike comets with large perihelion distances,} Monthly Notices of the Royal Astronomical Society, 472, 4634, \dodoi{10.1093/mnras/stx2157}

\bibitem[{J. Li(2023)Li}]{smplotlib}
Li, J. 2023, AstroJacobLi/smplotlib: v0.0.9, v0.0.9 Zenodo, \dodoi{10.5281/zenodo.8126529}

\bibitem[{B.~G. {Marsden} {et~al.}(1973){Marsden}, {Sekanina}, \& {Yeomans}}]{Marsden_1973}
{Marsden}, B.~G., {Sekanina}, Z., \& {Yeomans}, D.~K. 1973, \bibinfo{title}{{Comets and nongravitational forces. V},} \aj, 78, 211, \dodoi{10.1086/111402}

\bibitem[{K.~J. {Meech} {et~al.}(2017){Meech}, {Weryk}, {Micheli}, {Kleyna}, {Hainaut}, {Jedicke}, {Wainscoat}, {Chambers}, {Keane}, {Petric}, {Denneau}, {Magnier}, {Berger}, {Huber}, {Flewelling}, {Waters}, {Schunova-Lilly}, \& {Chastel}}]{Meech_1I_2017}
{Meech}, K.~J., {Weryk}, R., {Micheli}, M., {et~al.} 2017, \bibinfo{title}{{A brief visit from a red and extremely elongated interstellar asteroid},} \nat, 552, 378, \dodoi{10.1038/nature25020}

\bibitem[{M. {Micheli} {et~al.}(2018){Micheli}, {Farnocchia}, {Meech}, {Buie}, {Hainaut}, {Prialnik}, {Sch{\"o}rghofer}, {Weaver}, {Chodas}, {Kleyna}, {Weryk}, {Wainscoat}, {Ebeling}, {Keane}, {Chambers}, {Koschny}, \& {Petropoulos}}]{Oumuamua_nongrav_2018}
{Micheli}, M., {Farnocchia}, D., {Meech}, K.~J., {et~al.} 2018, \bibinfo{title}{{Non-gravitational acceleration in the trajectory of 1I/2017 U1 ('Oumuamua)},} \nat, 559, 223, \dodoi{10.1038/s41586-018-0254-4}

\bibitem[{C. {Opitom} {et~al.}(2019){Opitom}, {Fitzsimmons}, {Jehin}, {Moulane}, {Hainaut}, {Meech}, {Yang}, {Snodgrass}, {Micheli}, {Keane}, {Benkhaldoun}, \& {Kleyna}}]{Opitom_2I_2019}
{Opitom}, C., {Fitzsimmons}, A., {Jehin}, E., {et~al.} 2019, \bibinfo{title}{{2I/Borisov: A C$_{2}$-depleted interstellar comet},} \aap, 631, L8, \dodoi{10.1051/0004-6361/201936959}

\bibitem[{ {'Oumuamua ISSI Team} {et~al.}(2019){'Oumuamua ISSI Team}, {Bannister}, {Bhandare}, {Dybczy{\'n}ski}, {Fitzsimmons}, {Guilbert-Lepoutre}, {Jedicke}, {Knight}, {Meech}, {McNeill}, {Pfalzner}, {Raymond}, {Snodgrass}, {Trilling}, \& {Ye}}]{Oumuamua_ISSI_2019}
{'Oumuamua ISSI Team}, {Bannister}, M.~T., {Bhandare}, A., {et~al.} 2019, \bibinfo{title}{{The natural history of `Oumuamua},} Nature Astronomy, 3, 594, \dodoi{10.1038/s41550-019-0816-x}

\bibitem[{T. pandas~development team(2020)pandas~development team}]{pandas_2020}
pandas~development team, T. 2020, pandas-dev/pandas: Pandas, 2.2.3 Zenodo, \dodoi{10.5281/zenodo.3509134}

\bibitem[{S. Pilorz \& P. Jenniskens(2025)Pilorz \& Jenniskens}]{Pilorz_2025}
Pilorz, S., \& Jenniskens, P. 2025, \bibinfo{title}{Sun Close-Encounter model of Long-Period Comet and Meteoroid Orbit Stochastic Evolution,} Icarus, 437, 116559, \dodoi{https://doi.org/10.1016/j.icarus.2025.116559}

\bibitem[{G.~P. Prodan {et~al.}(2024)Prodan, Popescu, Licandro, Akhlaghi, de León, Tatsumi, Pastrav, Hibbert, Vǎduvescu, Simion, Pallé, Narita, Fukui, \& Murgas}]{prodan_2I_2024}
Prodan, G.~P., Popescu, M., Licandro, J., {et~al.} 2024, \bibinfo{title}{Pre-perihelion monitoring of interstellar comet 2I/Borisov,} Monthly Notices of the Royal Astronomical Society, 529, 3521, \dodoi{10.1093/mnras/stae539}

\bibitem[{H. {Rein} \& S.~F. {Liu}(2012){Rein} \& {Liu}}]{rebound}
{Rein}, H., \& {Liu}, S.~F. 2012, \bibinfo{title}{{REBOUND: an open-source multi-purpose N-body code for collisional dynamics},} \aap, 537, A128, \dodoi{10.1051/0004-6361/201118085}

\bibitem[{H. {Rein} \& D.~S. {Spiegel}(2015){Rein} \& {Spiegel}}]{reboundias15}
{Rein}, H., \& {Spiegel}, D.~S. 2015, \bibinfo{title}{{IAS15: a fast, adaptive, high-order integrator for gravitational dynamics, accurate to machine precision over a billion orbits},} \mnras, 446, 1424, \dodoi{10.1093/mnras/stu2164}

\bibitem[{R. {Sch{\"o}nrich} {et~al.}(2010){Sch{\"o}nrich}, {Binney}, \& {Dehnen}}]{Schonrich_2010}
{Sch{\"o}nrich}, R., {Binney}, J., \& {Dehnen}, W. 2010, \bibinfo{title}{{Local kinematics and the local standard of rest},} \mnras, 403, 1829, \dodoi{10.1111/j.1365-2966.2010.16253.x}

\bibitem[{D.~Z. {Seligman} {et~al.}(2025){Seligman}, {Micheli}, {Farnocchia}, {Denneau}, {Noonan}, {Hsieh}, {Santana-Ros}, {Tonry}, {Auchettl}, {Conversi}, {Devog{\`e}le}, {Faggioli}, {Feinstein}, {Fenucci}, {Ferrais}, {Frincke}, {Gillon}, {Hainaut}, {Hart}, {Hoffman}, {Holt}, {Hoogendam}, {Huber}, {Jehin}, {Kareta}, {Keane}, {Kelley}, {Lister}, {Mandt}, {Manfroid}, {Mar{\v{c}}eta}, {Meech}, {Amine Miftah}, {Morgan}, {Oca{\~n}a}, {Pe{\~n}a-Asensio}, {Shappee}, {Siverd}, {Taylor}, {Tucker}, {Wainscoat}, {Weryk}, {Wray}, {Yaginuma}, {Yang}, {Ye}, \& {Zhang}}]{3I_disc_2}
{Seligman}, D.~Z., {Micheli}, M., {Farnocchia}, D., {et~al.} 2025, \bibinfo{title}{{Discovery and Preliminary Characterization of a Third Interstellar Object: 3I/ATLAS},} \apjl, 989, L36, \dodoi{10.3847/2041-8213/adf49a}

\bibitem[{P.~M. {Shober} {et~al.}(2024){Shober}, {Tancredi}, {Vaubaillon}, {Devillepoix}, {Deam}, {Anghel}, {Sansom}, {Colas}, \& {Martino}}]{Shober_2024}
{Shober}, P.~M., {Tancredi}, G., {Vaubaillon}, J., {et~al.} 2024, \bibinfo{title}{{Comparing the dynamics of Jupiter-family Comets and comet-like fireballs},} \aap, 687, A181, \dodoi{10.1051/0004-6361/202449635}

\bibitem[{A. Sosa \& J.~A. Fernández(2011)Sosa \& Fernández}]{sosa_2011}
Sosa, A., \& Fernández, J.~A. 2011, \bibinfo{title}{Masses of long-period comets derived from non-gravitational effects – analysis of the computed results and the consistency and reliability of the non-gravitational parameters,} Monthly Notices of the Royal Astronomical Society, 416, 767, \dodoi{10.1111/j.1365-2966.2011.19111.x}

\bibitem[{D. Tamayo {et~al.}(2019)Tamayo, Rein, Shi, \& Hernandez}]{reboundx}
Tamayo, D., Rein, H., Shi, P., \& Hernandez, D.~M. 2019, \bibinfo{title}{REBOUNDx: a library for adding conservative and dissipative forces to otherwise symplectic N-body integrations,} Monthly Notices of the Royal Astronomical Society, 491, 2885, \dodoi{10.1093/mnras/stz2870}

\bibitem[{P. Virtanen {et~al.}(2020)Virtanen, Gommers, Oliphant, Haberland, Reddy, Cournapeau, Burovski, Peterson, Weckesser, Bright, {van der Walt}, Brett, Wilson, Millman, Mayorov, Nelson, Jones, Kern, Larson, Carey, Polat, Feng, Moore, {VanderPlas}, Laxalde, Perktold, Cimrman, Henriksen, Quintero, Harris, Archibald, Ribeiro, Pedregosa, {van Mulbregt}, \& {SciPy 1.0 Contributors}}]{2020SciPy-NMeth}
Virtanen, P., Gommers, R., Oliphant, T.~E., {et~al.} 2020, \bibinfo{title}{{{SciPy} 1.0: Fundamental Algorithms for Scientific Computing in Python},} Nature Methods, 17, 261, \dodoi{10.1038/s41592-019-0686-2}

\bibitem[{Q.-Z. {Ye} {et~al.}(2017){Ye}, {Zhang}, {Kelley}, \& {Brown}}]{Quan_1I_2017}
{Ye}, Q.-Z., {Zhang}, Q., {Kelley}, M. S.~P., \& {Brown}, P.~G. 2017, \bibinfo{title}{{1I/2017 U1 ({\textquoteleft}Oumuamua) is Hot: Imaging, Spectroscopy, and Search of Meteor Activity},} \apjl, 851, L5, \dodoi{10.3847/2041-8213/aa9a34}

\end{thebibliography}
\bibliographystyle{aasjournalv7}



\end{document}